\documentclass[11pt, superscriptaddress]{revtex4-2}
\usepackage{graphics,epsfig}
 \usepackage{hyperref}
\usepackage{epstopdf}
\usepackage{graphicx}
 
\usepackage{dcolumn}
\usepackage{amsmath}
\newcommand{\be}{\begin{equation}}
\newcommand{\ee}{\end{equation}}
\newcommand{\bea}{\begin{eqnarray}}
\newcommand{\eea}{\end{eqnarray}}
\newcommand{\nn}{\nonumber}

\begin{document}

\title{Black hole solution and thermal properties in $4D$ $AdS$ Gauss-Bonnet massive gravity }

\author{Sudhaker Upadhyay}
\email{sudhakerupadhyay@gmail.com}
\affiliation{Department of Physics, K. L. S. College, Nawada, Bihar 805110, India}
\affiliation{Department of Physics, Magadh University, Bodh Gaya,
 Bihar  824234, India}
\affiliation{Inter-University Centre for Astronomy and Astrophysics (IUCAA), Pune, \\Maharashtra 411007, India}
\affiliation{School of Physics, Damghan University, P.O. Box 3671641167, \\ Damghan, Iran}
 \author{Dharm Veer Singh}
\email{veerdsingh@gmail.com}
\affiliation{Department of Physics, Institute of Applied Science and Humanities, GLA University, Mathura, 281406 India.}

\begin{abstract}
 We consider an  Einstein-Gauss-Bonnet massive gravity model in $4D$ $AdS$ spacetime to obtain a possible black hole solution and discuss the horizon structure of this black hole. The real roots of the vanishing metric function lead to various types of horizons. Furthermore, we derive various  thermodynamic quantities, thus insuring the validity of the first-law of thermodynamics and the Smarr relation.   The thermodynamic  quantities are modified in the presence of the massive gravity parameter and also discuss the stability  of the system from the heat capacity. Black hole space times can not only  possess  standard thermodynamics, but also possess phase structures when the cosmological constant is treated as the thermodynamic  pressure.   The effects of massive parameter and GB parameter  on phase transition are also discussed.   We also examine the phase structure through Maxwell's equal area law. The first order phase transition occurs only when pressure is lower than its critical value. At critical pressure, the first order phase transition terminates, and the second order phase transition occurs.  No phase transition occurs when pressure is greater than its critical value. 
\end{abstract} 
\maketitle
 
\section{Introduction}
The complete understanding of the universe from the principles of general relativity (GR) is one of the challenging tasks for present day science as GR is not theoretically complete. The explanation of dark matter and dark energy present in the Universe is one of them. Various efforts by various people were made to complete the GR by modifying the Einstein theory of gravity, but the conclusive theory is still missing. To be more precise, some of the examples of modified theories of gravity are scalar-tensor theories \cite{1,2,3,4,5,6,7}, Lovelock gravity \cite{8,9,10} and brane world cosmology \cite{11,12,13}. Here, we place emphasis on the higher derivative of gravity known as the Lovelock theory, which extends Einstein's theory to higher dimensions. 
 
The Lovelock's theorem \cite{8,9} states that Einstein's GR with the cosmological constant is the unique theory of gravity with four-dimensional spacetime, diffeomorphism invariance, metricity, and second-order equations of motion.  Lovelock's gravity \cite{8}, among the other gravity theories with higher derivative curvature terms, has some special features.  Gauss-Bonnet (GB) gravity is a particular case of Lovelock gravity in higher dimensions \cite{lan}.  In Ref. \cite{cai}, it is shown that for the GB black holes in $AdS$ spacetime, there exists a minimal horizon radius below which the Hawking-Page phase transition will not occur. Recently,  the absorption cross section of planar scalar massless waves impinging on   solutions of the   $4D$  Einstein-Gauss-Bonnet (EGB) theory of gravity \cite{har}. The solution depends also on the mass of the black hole and the GB constant coupling.
 
One of the possibilities to modify GR by considering a massive graviton also attracts the attention of various people.  For instance, a ghost-free theory with massive gravitons  is developed in Ref. \cite{14}. In curved spacetime,  this causes to the presence of   ghost instabilities \cite{15}. The   black hole  solutions  in the presence of massive gravity  with their effects   on the geometrical structure is also studied  \cite{16,17}.  Hendi et al. \cite{Hendi:2015pda}  investigated charged black hole solutions in GB massive gravity and their thermodynamics in $d$
 dimensions. 
 
  In fact, the GB term in $4D$ is a total derivative and thus   does not contribute to the dynamics. 
However, the GB term contributes to the dynamics of the gravitational field only for $ d>4$.  When the GB term in $4D $ is coupled to a matter field, it contributes to dynamics.  In an interesting work, Glavan and Lin   recently found    EGB gravity in four dimensions by rescaling the GB coupling constant $\alpha \to \alpha/D-4$, where the GB term   contributes to the local dynamics \cite{gla}. Indeed, taking the $D\to 4 $ limit either breaks part of the diffeomorphism. However it is possible to construct a Lorents violating theory by breaking the diffeomorphism and it is invariant only under $3D$ spatial diffeomorphism as the same number of degree of freedom as general relativity \cite{Ao1,Ao2}. One can find a consistent  $4D$ EGB gravity that is covariant under spatial diffeomorphism  only but not under time  diffeomorphism. The consistent theory validates some results of the solution proposed by  Glavin and Lin \cite{gla}. The generlization of other spherically symmetric black hole solution has been discussed in \cite{fran,Hennigar:2020lsl,Singh:2020mty,Singh:2020nwo,Singh:2020xju,Singh:2021xbk,Singh21,Singh20,Godani:2022jwz,hendi19,hendi20,hendi18,hendi17,hendi16,hendi191,hendi2017,Wei:2020poh}. So, there is enough motivation to generalize the result by studying the black hole solution for the massive EGB  gravity in four dimensions.  This is an objective of the present work. 

In this paper, we intend to generalize EGB action by adding a massive term and obtain an exact  
solution of the  massive EGB gravity in $4D$ $AdS$ spacetime. We will discuss horizon structure and obtain various types of horizons. Furthermore, we analyse thermal properties and the validity of the first-law of thermodynamics. We also discuss the role of graviton mass and the GB parameter in the stability of this black hole solution.  Finally, we shed light on the graviton mass and GB parameter dependence of phase transition and critical points.

The remaining parts of the paper are organized as follows. In  Sec. \ref{s2}, we obtain a
black hole solution and discuss the horizons of the  EGB massive gravity in $4D$ $AdS$ space.
In Sec. \ref{s3}, we provide the thermodynamics of the system. The stability of the black hole
is analyzed in section \ref{s4}. The phase transition is presented in section \ref{s5}.
The paper is summarized with concluding remarks in the last section. 
\section{Black hole solution in $4D$ EGB massive gravity}\label{s2}
Let us begin by writing the action for the  massive Einstein-Gauss-Bonnet (EGB) gravity as following:
\begin{eqnarray}
S&=&\frac{1}{2}\int d^{d}x\sqrt{-g}\left[ {\cal R}-2\Lambda + {\alpha}  {\cal L_{GB}}+m^2\sum_{i}c_i{\cal U}_i(g,h) \right],
\label{action}
\end{eqnarray}
where ${\cal R}$ is the curvature scalar,  $\alpha$  is a Gauss-Bonnet (GB) coupling constant
 with dimension [length]$^2$, $ \mathcal{L_{GB}}:=R_{\mu \nu \gamma \delta }R^{\mu \nu \gamma \delta}-4R_{\mu \nu }R^{\mu \nu }+R^{2}$ is the GB Lagrangian density. Additionally,  $m$ represents massive gravity parameter related to the graviton mass, $h$ refers to the fixed symmetric tensor, $c_{i}$  are free  constants and $\mathcal{U}_{i}(g,f)$ refer to the symmetric polynomials of the eigenvalues of   matrix $ 
\mathcal{K}_{\nu }^{\mu }=\sqrt{g^{\mu \alpha }h_{\alpha \nu }}$ \cite{Singh20, Singh21}. The components of $\mathcal{U}_{i}(g,h)$ is expressed by  
\begin{eqnarray}
\mathcal{U}_{1} &=&\left[ \mathcal{K}\right] ,\nn \\
\mathcal{U}_{2} &=&\left[ \mathcal{K}\right] ^{2}-\left[ \mathcal{K}^{2}
\right] ,\nn \\
\mathcal{U}_{3} &=&\left[ \mathcal{K}\right] ^{3}-3\left[ \mathcal{K}\right] 
\left[ \mathcal{K}^{2}\right] +2\left[ \mathcal{K}^{3}\right] ,\nn \\
\mathcal{U}_{4} &=&\left[ \mathcal{K}\right] ^{4}-6\left[ \mathcal{K}^{2}
\right] \left[ \mathcal{K}\right] ^{2}+8\left[ \mathcal{K}^{3}\right] \left[
\mathcal{K}\right] +3\left[ \mathcal{K}^{2}\right] ^{2}-6\left[ \mathcal{K}
^{4}\right],\label{u}
\end{eqnarray}
where the $[\ ]$  represents the trace of matrix ${\cal K}_{\nu}^{\mu}$. By  varying the action (\ref{action}) with respect to metric tensor $g_{\mu\nu}$,  the equation of motion is calculated 
as
\begin{eqnarray}
G_{\mu \nu }+\Lambda g_{\mu \nu }+H_{\mu \nu }+m^{2}\chi _{\mu \nu }=0,
\label{Field equation}
\end{eqnarray}
where  Einstein tensor ($G_{\mu \nu}$), Lanczos tensor ($H_{\mu \nu} $)  and  massive tensor ($ \chi_{ \mu \nu} $) have following explicit expressions, respectively: 
\begin{eqnarray}
G_{\mu\nu}&=&R_{\mu\nu}-\frac{1}{2}g_{\mu\nu}R,\\
H_{\mu \nu }& =&-\frac{\alpha }{2}\left[ 8R^{\rho \sigma }R_{\mu \rho \nu
\sigma }-4R_{\mu }^{\rho \sigma \lambda }R_{\nu \rho \sigma \lambda
}-4RR_{\mu \nu }+8R_{\mu \lambda }R_{\nu }^{\lambda }\right.  \nn  \\
&+&\,\,\left. g_{\mu \nu }\left( R_{\mu \nu \gamma \delta }R^{\mu \nu \gamma
\delta }-4R_{\mu \nu }R^{\mu \nu }+R^{2}\right) \right],\\
\chi _{\mu \nu }& =&-\frac{c_{1}}{2}\left( \mathcal{U}_{1}g_{\mu \nu }-
\mathcal{K}_{\mu \nu }\right) -\frac{c_{2}}{2}\left( \mathcal{U}_{2}g_{\mu
\nu }-2\mathcal{U}_{1}\mathcal{K}_{\mu \nu }+2\mathcal{K}_{\mu \nu
}^{2}\right) \nn  \\
&-&\,\,\frac{c_{3}}{2}(\mathcal{U}_{3}g_{\mu \nu }-3\mathcal{U}_{2}
\mathcal{K}_{\mu \nu } +6\mathcal{U}_{1}\mathcal{K}_{\mu \nu }^{2}-6\mathcal{K}_{\mu \nu }^{3})\nn \\
&-&\,\,\frac{c_{4}}{2}(\mathcal{U}_{4}g_{\mu \nu }-4\mathcal{U}_{3}\mathcal{K}_{\mu
\nu }+12\mathcal{U}_{2}\mathcal{K}_{\mu \nu }^{2}-24\mathcal{U}_{1}\mathcal{K
}_{\mu \nu }^{3}+24\mathcal{K}_{\mu \nu }^{4}). 
\end{eqnarray}
To have a   static spherically symmetric black hole solution,
The spherically symmetric solution   after re-scaling
the coupling constant by $\alpha/(D- 4)$, in the limit $D \rightarrow 4$, takes the line element as follows:
\begin{equation}
ds^2 = -f(r)dt^2+\frac{1}{f(r)} dr^2 + r^2 d\Omega_{d-2},
\label{metric}
\end{equation}
where $  d\Omega_{D-2}$ is the metric of a $(d-2)$-dimensional constant curvature space. Now, we make ansatz for reference metric as
\begin{equation}
h_{\mu \nu }=\mbox{diag}(0,0,c^{2}, c^2 \sin^2\theta),
  \label{f11}
\end{equation}
where  $c$ is a positive constant. 
Using the metric ansatz (\ref{f11}), the components of $\mathcal {U}_i$ (\ref{u}) identify to \cite{Hendi:2015pda}
\begin{eqnarray}
&&\mathcal {U}_1=\frac{(d-2)c}{r},\qquad \qquad \qquad \mathcal {U}_3=\frac{(d-2)(d-3)(d-4)c^3}{r^3},\\
&&\mathcal {U}_2=\frac{(d-2)(d-3) c^2}{r^2}\qquad\qquad \mathcal {U}_4=\frac{(d-2)(d-3)(d-4)(d-5) c^4}{r^4}.
\end{eqnarray}
The  $(r,r)$ components of Einstein field equation  Eq. (\ref{Field equation}) in the limit $d\to 4$ gives
\begin{eqnarray}
&&r^5-2r^3\alpha(f-1)f'+r^4(f-1)+  r^2\alpha (f-1)^2-\Lambda r^2-m^2(cc_1r +c^2c_2)=0,
\label{rr}
\end{eqnarray}
where prime denotes derivative with respect to $r$. 
 We   obtain the following solution by solving the Eq. (\ref{rr}) 
\begin{eqnarray}
f_{\pm}(r)=1+\frac{r^2}{2\alpha}\left(1\pm\sqrt{1+4\alpha\left(\frac{2M}{r^3}-\frac{1}{l^2}-\frac{m^2}{2r^2}\left(cc_1 r+2c^2c_2\right)\right)}\,\right),
\label{sole}
\end{eqnarray}
where $M$ is an integration constant  related to the total mass of the black hole. Here cosmological constant is expressed in terms of scale length factor $l$ which, in general, takes value $(-3/l^2)$   for  $AdS$  solutions.
 This is an exact solution of the field equation (3).
 In the  solution (\ref{sole}), $M$ refers to the integration constant related to the mass of the black hole, $\alpha$ refers to the Gauss-Bonnet coupling, and {  dimensionless quantity $m$ is related to the mass of  graviton}. For the further analysis, it is convenient to work in dimensionless parameters:
\begin{equation}
{\tilde{r}}=\frac{r}{l}, \quad{\tilde M}=\frac{M}{l},\quad {\tilde \alpha}=\frac{\alpha}{l^2},\quad \text{and} \quad {\tilde c_1}=\frac{c_1}{l}.
\end{equation}
The solution (\ref{sole}) can be expressed in terms of the above parameters as
\begin{eqnarray}
f_{\pm}({\tilde r})=1+\frac{{\tilde r}^2}{2 {\tilde \alpha}}\left(1\pm\sqrt{1+4{\tilde\alpha}\left(\frac{2{\tilde M}}{{\tilde{r}^3}}-\frac{{ m}^2}{2{\tilde r}^2}\left(c\tilde c_1{\tilde r}+2c^2c_2\right)\right)-1}\,\right),
\label{sole11}
\end{eqnarray}

 The solution (\ref{sole11})  behaves asymptotically as
\begin{eqnarray}
&&f_{-}=1-\frac{2{\tilde M}}{\tilde r}+{\tilde r^2}+\frac{m^2}{2}(c\tilde c_1r +c^2c_2)+\frac{\tilde r^2}{2\tilde \alpha}+{{\cal O}\left(\frac{1}{r^3}\right)},\\
&&f_+= 1+\frac{2\tilde M}{\tilde r}-{\tilde r^2}-\frac{ m^2}{2}(c \tilde c_1{\tilde r} +c^2c_2)+\frac{\tilde r^2}{2\tilde \alpha}+{{\cal O}\left(\frac{1}{\tilde r^3}\right)}.
\end{eqnarray}
The $-$ve branch corresponds to the $4D $ $AdS $ EGB massive black hole, whereas the $+ $ve branch does not lead to a physically meaningful solution because the positive sign in the mass term indicates graviton instabilities, so we will stick to the $-$ve branch of (\ref{sole11}). Furthermore if we take the limit $\tilde r \to \infty$ ($\tilde M = 0$), the $-$ve  branch of the solution (\ref{sole11}) is asymptotically flat whereas the $+$ve branch of the solution (\ref{sole11}) is asymptotically $dS$ ($AdS$) depending on the sign of $\tilde\alpha$.  

 In the massless graviton limit ($m=0$), this solution reduces to
\begin{eqnarray}
f_{\pm}(\tilde r)=1+\frac{\tilde r^2}{2\tilde \alpha}\left(1\pm\sqrt{1+\frac{8\tilde M\tilde \alpha}{\tilde r^3}-{ 4\tilde\alpha}}\,\right).
\label{sole1}
\end{eqnarray}
This  black hole solution (\ref{sole1}) coincides  with  the  solution given by Glavan and Lin in Ref. \cite{gla} when $m=0$ and matches with $AdS$ Schwarzschild massive  black hole solution when $\alpha =0$.

\begin{figure}[h]
\begin{tabular}{c c c c}
\includegraphics[width=.50\linewidth]{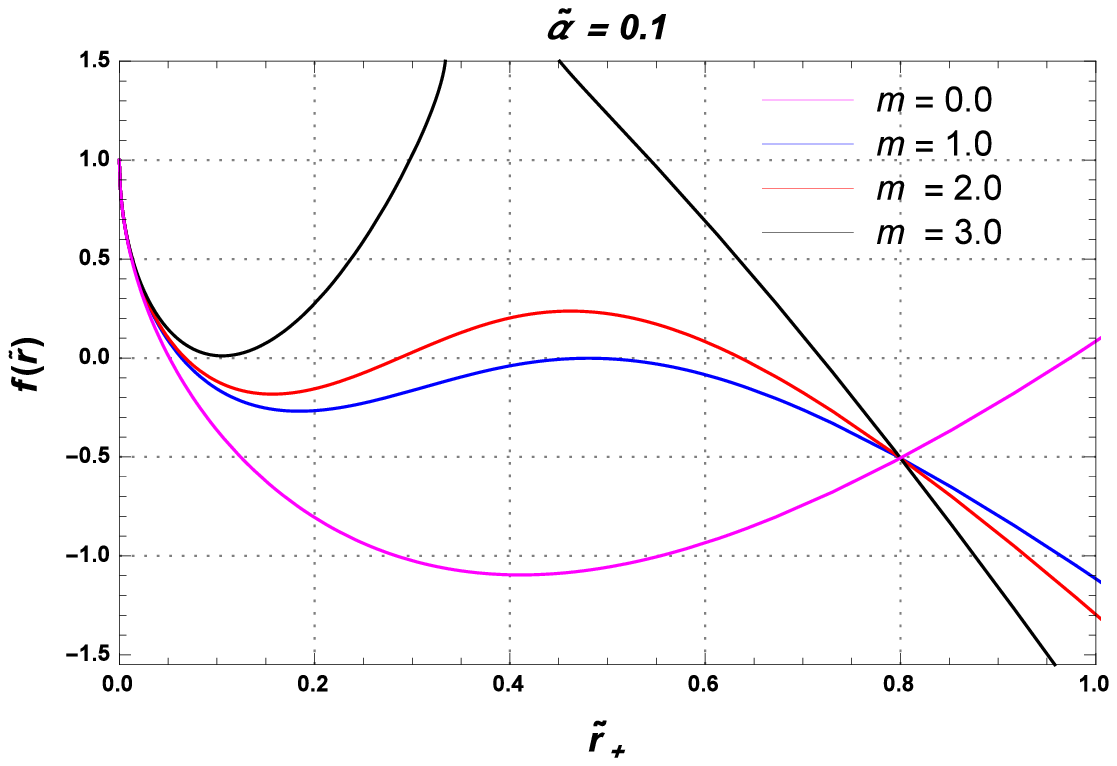}
\includegraphics[width=.50\linewidth]{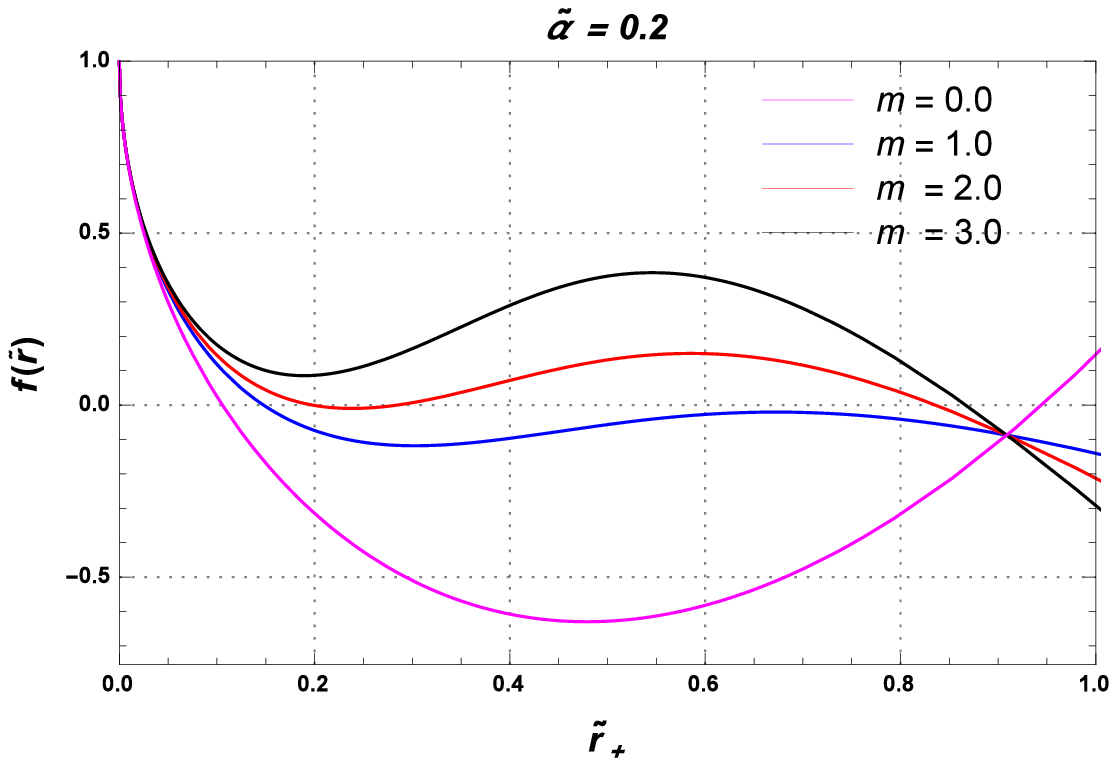} 
\end{tabular}
\caption{Metric function plot $f(r)$  as a function of horizon radius with the variation of massive gravity parameter $({ m})$ for GB parameter ${\tilde \alpha}=0.1$ and ${\tilde \alpha}=0.2$. Here,  $c=1,\tilde c_1=-1, c_2=1$. }
\label{fig1} 
\end{figure}

In order to study the   horizon structure,  we plot  these thermodynamics quantities in Fig. \ref{fig1} as the function of horizon radius with different value of massive gravity parameter with specific GB parameter $\tilde \alpha=0.1$ and $\tilde \alpha=0.2$.  The Fig. \ref{fig1} describes  the horizon structure of the  $4D$ EGB massive black hole. The numerical analysis of $f (r_+) = 0$  reveals that it is possible to find non-vanishing value of   $\alpha$, $m$ and  cosmological  constant ($\Lambda$)  for which metric function  is minimum.   The metric function $f (r_+ ) = 0$    gives three real roots $r_+$ (Cauchy horizon), $r_-$ (event horizon) and $r_c$ (cosmological horizon).
	\begin{table}[h]
		\begin{center}
			\begin{tabular}{l l r l| r l r l r}
				\hline
				\hline
				\multicolumn{1}{c}{ }&\multicolumn{1}{c}{ $\tilde \alpha=0.1$  }&\multicolumn{1}{c}{}&\multicolumn{1}{c|}{ \,\,\,\,\,\, }&\multicolumn{1}{c}{ }&\multicolumn{1}{c}{}&\multicolumn{1}{c}{ $\tilde\alpha=0.2$ }&\multicolumn{1}{c}{}\,\,\,\,\,\,\\
				\hline
				\multicolumn{1}{c}{ \it{$m$}} & \multicolumn{1}{c}{ $r_-$ } & \multicolumn{1}{c}{ $r_+$ }& \multicolumn{1}{c|}{$\delta$}&\multicolumn{1}{c}{\it{$m$}}& \multicolumn{1}{c}{$r_-$} &\multicolumn{1}{c}{$r_+$} &\multicolumn{1}{c}{$\delta$}   \\
				\hline
				\,\,\, 0\,\,& \,\,0.1344\,\, &\,\, 1.934\,\,& \,\,1.799\,\,&$0$&\,\, 0.0963\,\,&\,\,1.865\,\,&\,\,1.768\,\,
				\\
				\
				 1.50\,\, & \,\,0.3685\,\, &\,\, 1.715\,\,& \,\,1.346\,\,&$1.50$&\,\,0.3716\,\,&\,\,1.5\,\,&\,\,1.128\,\,
				\\
				 \,\,\,2.09\,\, &  \,\,0.3539\,\,  &\,\,1.429\,\,&\,\,0.5996\,\,&1.73 &\,\,0.5023\,\,&\,\,1.328\,\,&\,\,0.8257\,\,
	\\
				 \,\,\,2.50\,\, &  \,\,0.3539\,\,  &\,\,1.429\,\,&\,\,0.5996\,\,&1.90 &\,\,0.5023\,\,&\,\,1.328\,\,&\,\,0.8257\,\,
				\\
	
				\hline
				\hline
			\end{tabular}
		\end{center}
		\caption{Cauchy ($\tilde r_{-}$) and event ($\tilde r_{+}$) horizons, and $\delta=\tilde r_+-\tilde r_-$ for  the $4D$  AdS EGB massive black hole  for GB parameter ${\tilde \alpha}=0.1$ and ${\tilde \alpha}=0.2$. Here,  $c=1,\tilde c_1=-1, c_2=1$.}
		\label{tab:temp1}
		\label{f3}
	\end{table}

From Fig. \ref{fig1},  it is clear  that  the size of the black hole increases with a decrease in the GB parameter. The black hole has three (Cauchy, event, and cosmological) horizons at $ m> 2.09$ with $\tilde\alpha=0.1$. However, the event  and cosmological horizons  are possible  for $ m<2.09$.
The  horizon is a decreasing function of massive parameter ($ m$) (Fig. \ref{fig1}  right panel), an increasing function of massive gravity parameter and finally independent of GB parameter (Fig. \ref{fig1} right panel). It should be pointed out that interestingly, increasing and decreasing mentioned parameters, will lead to formation of the different sized black hole. 

\section{Thermodynamics}\label{s3}
In this section, we study the thermodynamic properties of $4D$ $AdS$ Gauss-Bonnet massive black hole.
In particular, these are   mass $(\tilde M_+)$, temperature $(\tilde T_+)$, entropy $(\tilde S_+)$ and free energy $(\tilde G_+)$. The mass of the black hole is derived by setting $f_-(\tilde r)|_{\tilde r=\tilde r_+}=0$ as
\begin{eqnarray}
 \tilde M_+=\frac{1}{2\tilde r_+}\left[\tilde r_+^2+ \tilde \alpha +{\tilde r_+^4}+ { m^2\tilde r_+^2} \left( \frac{c \tilde c_1\tilde r_+}{2} +2c^2c_2 
 \right)\right]. \label{m}
\end{eqnarray}
The Hawking temperature for this black hole solution can be estimated from the well-known area-law
\begin{equation}
\tilde T_+=\frac{1}{4\pi}f'(\tilde r)|_{\tilde r=\tilde r_+},
\end{equation} 
as follows 
\begin{eqnarray}
\tilde T_+=T_+ l=\frac{1}{4\pi \tilde r_+(r_+^2+2\tilde \alpha)}\left[\tilde r_+^2-\tilde \alpha+{3\tilde r_+^4}+m^2\tilde r_+^2(c^2c_2+c\tilde c_1 \tilde r)\right].
\label{t}
\end{eqnarray}
\begin{figure}[h]
\begin{tabular}{c c c c} 
\includegraphics[width=.50\linewidth]{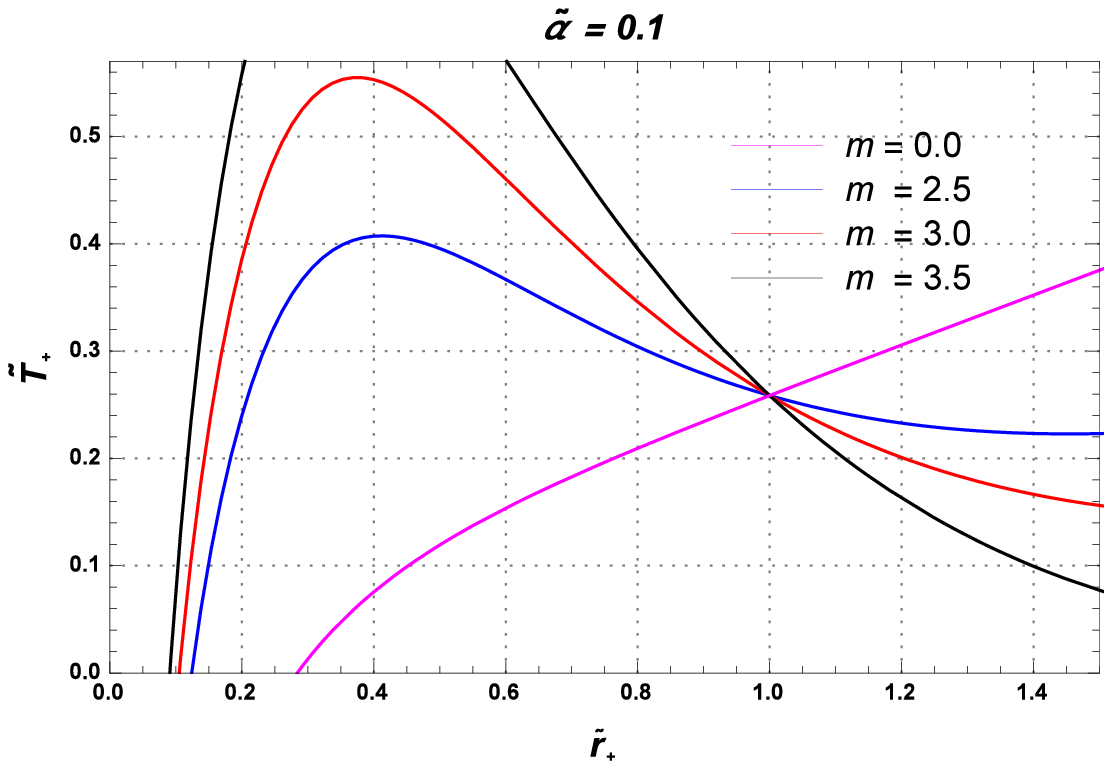}
\includegraphics[width=.50\linewidth]{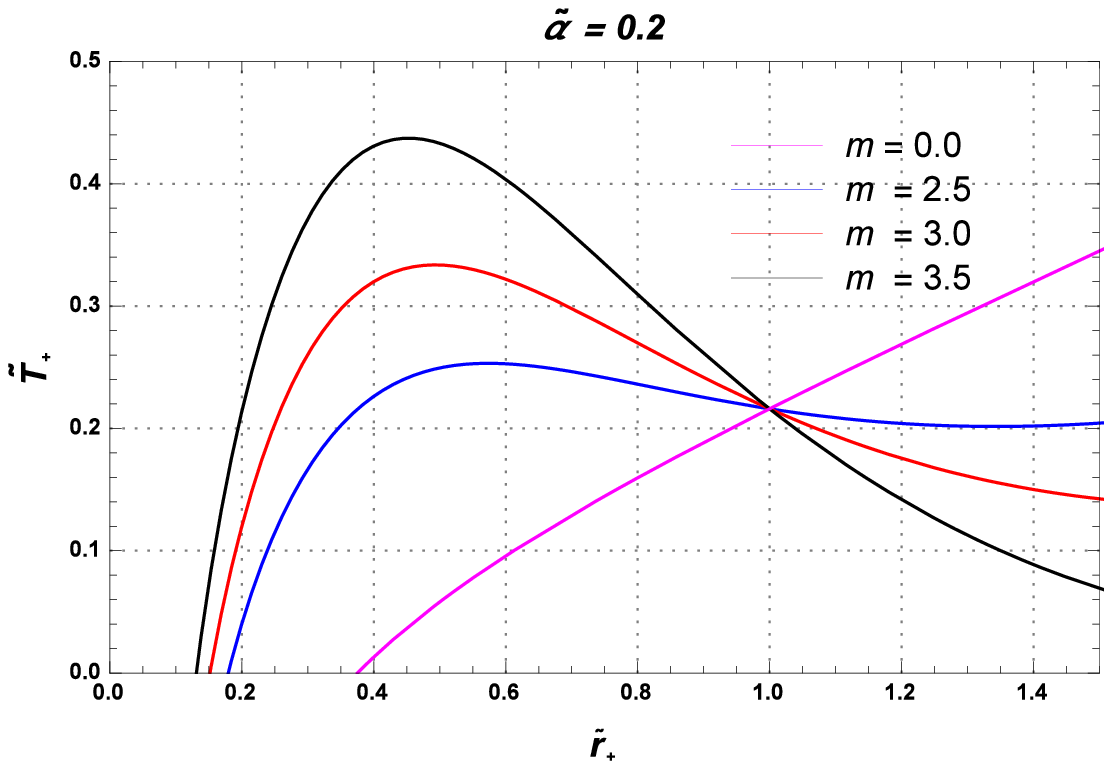} 
\end{tabular}
\caption{Temperature plot $(\tilde T_+)$ as a function of horizon radius with the variation of massive gravity parameter $( m)$ for GB parameter $\tilde \alpha=0.1$ and $\tilde \alpha=0.2$. Here, $c=1,\tilde c_1=-1, c_2=1$. }
\label{fig2} 
\end{figure}
It is evident Fig. \ref{fig2}  that the temperature is negative when the two (Cauchy and event) horizons    merge, therefore, in this region, the  obtained solutions are not physical. 

\noindent The thermodynamical quantities must follow  the first law of thermodynamics
\be
d\tilde M_+=\tilde Td\tilde S_+.
\ee
This leads to explicit expression for entropy $\tilde S_+=\int \frac {d\tilde M_+}{\tilde T_+} d\tilde r_+$ as \cite{Wei:2020poh,HosseiniMansoori:2020yfj}
\begin{eqnarray}
\tilde S_+=\pi \tilde r_+^2 +4\pi\tilde \alpha \log [\tilde r]+\tilde S_0,
\end{eqnarray} 
where $\tilde S_0$ is an integration  constant.  

It is worth mentioning that the definition of first-law of thermodynamics may have additional terms also in the extended phase space. In order to check the validity of these terms one should check the first-law in the non-differential form  called as the  Smarr relation
\be
\tilde M_+=2\tilde T_+ \tilde S_++2\tilde V_+ \tilde P-\tilde A\tilde \alpha-{\cal \tilde C}\tilde c_1,\label{me}
\ee
where
\bea
&&\tilde P=-\frac{\Lambda}{8\pi},\qquad\qquad\qquad\qquad\quad \tilde V_+=\left(\frac{\partial \tilde M_+}{\partial \tilde P}\right)_{\tilde S_+,\tilde\alpha, \tilde c_1},\\
&&{\cal \tilde C}=\left(\frac{\partial \tilde M_+}{\partial \tilde c_1}\right)_{\tilde S_+,\tilde \alpha, \tilde P},\qquad\qquad\qquad \tilde A=\left(\frac{\partial \tilde M_+}{\partial \tilde \alpha}\right)_{\tilde S_+,\tilde P,\tilde c_1}.
\eea
From all the above additional  thermodynamic quantities, it is easy to verify the complete form of the first-law of thermodynamics in an extended space
\be
d\tilde M_+=\tilde T_+d\tilde S_+-\tilde V_+d\tilde P+\tilde Ad\tilde \alpha+{\cal \tilde C}dc_1.
\label{fl1}
\ee
To understand the global stability of the black hole,  the behavior of the Gibbs free energy $\tilde G_+$ is important to analyzed. The Gibbs free energy,  from  the definition $\tilde G_+=\tilde M_+-\tilde T_+\tilde S_+$, is derived by
\bea
\tilde G_+&=&\frac{1}{2\tilde r_+}\left[\tilde r_+^2+2\tilde \alpha+{\tilde r_+^3}+m^2\left(\frac{c \tilde c_1\tilde  r_+^2}{2}+c^2c_2 \tilde r_+\right)\right]\nn\\&-& \frac{(\tilde  r_+^2+4\alpha \log[\frac{\tilde r}{\tilde r_0}]) }{4 \tilde r_+( \tilde r_+^2+2\tilde \alpha)}\left[\tilde r_+^2-\tilde \alpha+{3\tilde r_+^4}+ m^2\tilde r_+^2(c^2c_2+c\tilde c_1\tilde r)\right].\label{gibbs}
\eea
\begin{figure}[h]
\begin{tabular}{c c c c} 
\includegraphics[width=.50\linewidth]{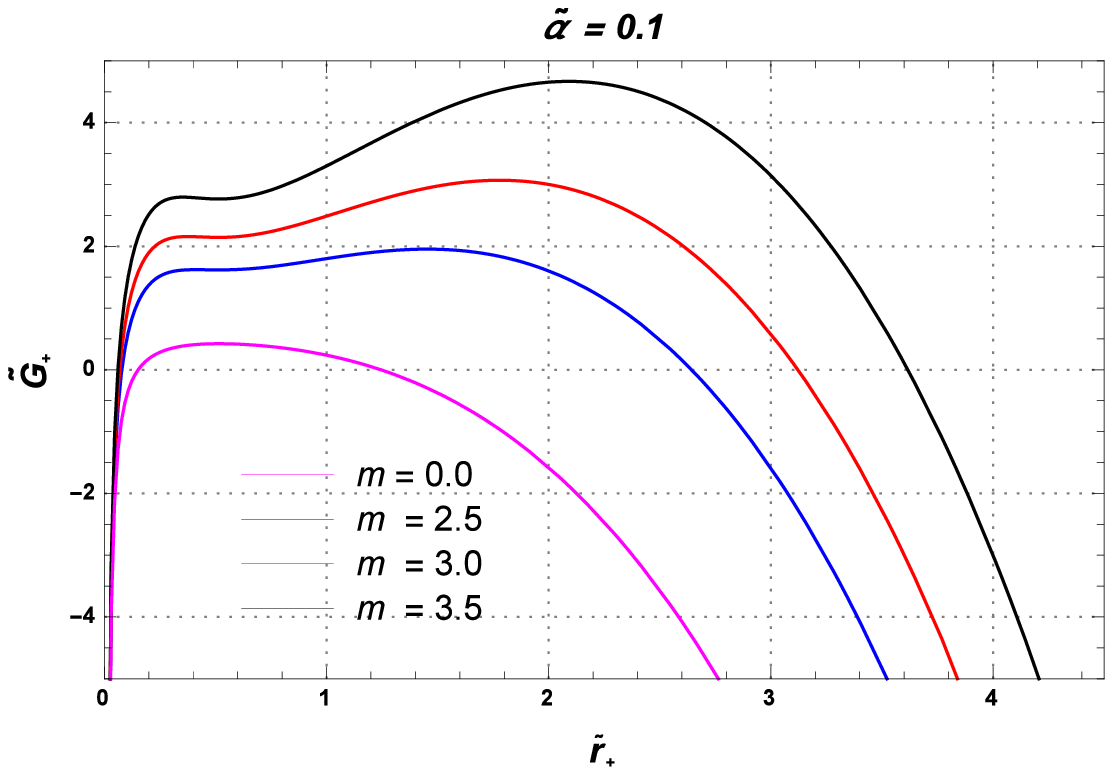}
\includegraphics[width=.50\linewidth]{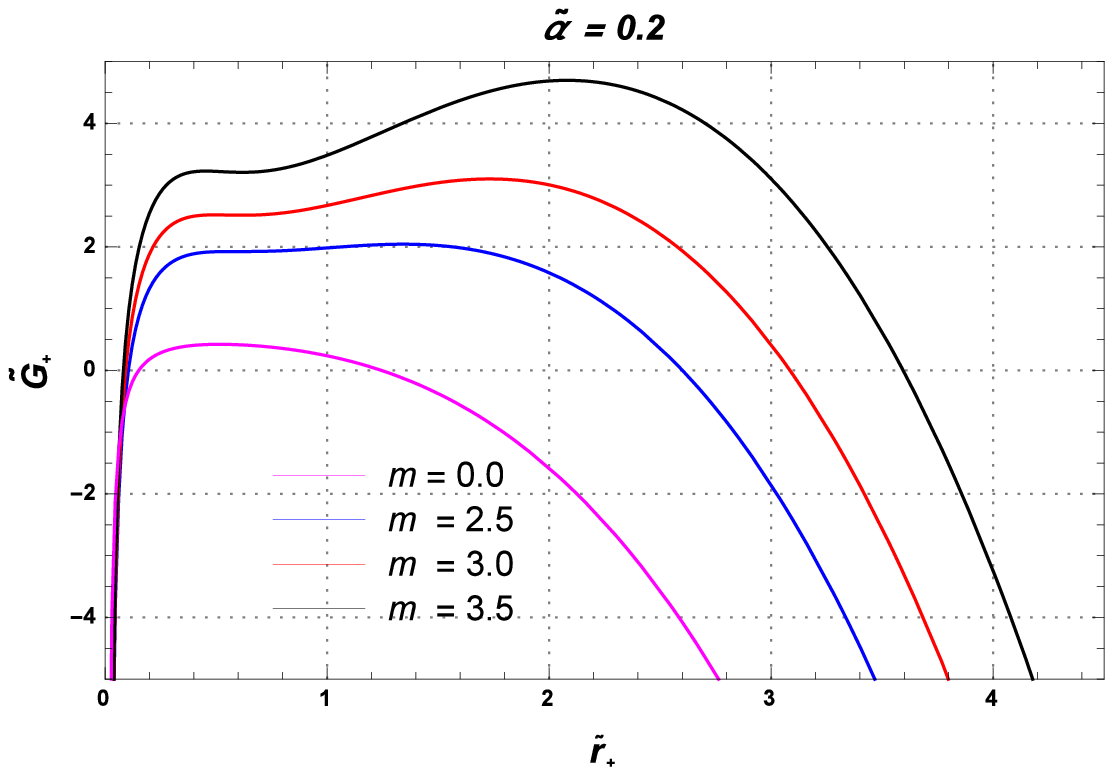}
\end{tabular}
\caption{Plot of   free energy $(\tilde G_+) $ as a function of horizon radius with the variation of massive gravity parameter $(m)$ for GB parameter $\tilde \alpha=0.1$ and $\tilde \alpha=0.2$. Here, $c=1,c_1=-1, c_2=1$. }
\label{fig3} 
\end{figure}
To analyze Gibbs free energy, we plot figure \ref{fig3} for different values of massive gravity parameter   and  GB parameter.   Here, we see that Gibbs free energy increases with  GB parameter. However, with increasing $m$ the Gibbs free energy first increases and then starts falling  after a critical point.  

\section{Heat Capacitity and thermodynamics stability} \label{s4}
Now, we study the heat capacity of the obtained black hole solution (\ref{sole}) in the context of canonical ensemble by calculating the heat capacity. The heat capacity is given by the following relation:
\begin{equation}
\tilde C_+=\left(\frac{\partial \tilde  M_+}{\partial \tilde T_+}\right)_{\tilde \alpha}=  \left(\frac{\tilde \partial M_+}{\tilde \partial r_+}\right)_{\tilde \alpha}\left(\frac{\partial \tilde r_+}{\partial \tilde T_+}\right)_{\tilde \alpha}.
\label{sh}
\end{equation}
Substituting the value of mass  from   (\ref{m}) and  temperature from   (\ref{t}) in Eq. (\ref{sh}), we get
\be
\tilde C_+=\frac{{2  \pi \tilde r^2 (\tilde r^2 + 2\tilde \alpha)^2}\left (1 + c^2 c_2 \tilde  m^2 + c\tilde c_1  m^2 r + {3\tilde r^2} - \frac{\tilde \alpha}{\tilde r^2}\right)}{3 (\tilde r^6 + 6 \tilde r^4 \tilde \alpha) + 
  [-(1 + c^2 c_2  m^2)\tilde r^4 + (5 + 2 c^2 c_2  m^2)\tilde r^2\tilde \alpha + 
    4 c\tilde c_1 m^2 \tilde r^3 \tilde \alpha + 2 \tilde \alpha^2]}.\label{sh1}
\ee
\begin{figure}[h]
\begin{tabular}{c c c c} 
 \includegraphics[width=.50\linewidth]{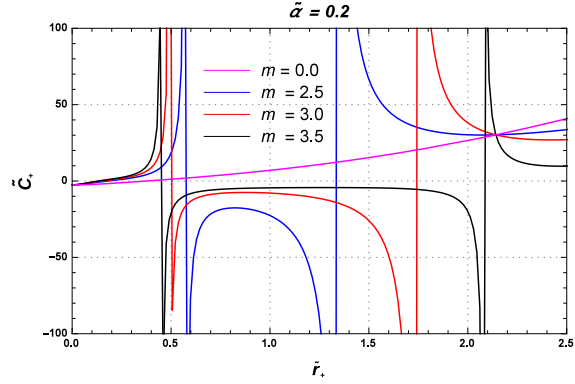}
\end{tabular}
\caption{Plot of  heat capacity ($\tilde C_+$)  as a function of horizon radius with the variation of massive gravity parameter $( m)$ for GB parameter  $\tilde \alpha=0.2$ with fixed value of $c=1,\tilde c_1=-1, c_2=1$. }
\label{fig4} 
\end{figure}
In order to study the thermodynamical behavior of the system 
  stability and phase transition points, we plot  this thermodynamics quantity in Fig. \ref{fig4}.
For Eq. (\ref{sh1}),   the heat capacity will have divergences at extremum points. 

 We know that roots and divergences
present in the heat capacity   represent phase transition points of the system. Additionally, we should mention that unstable systems    undergo a phase transition and becomes stable.  Corresponding to the root there exists a
phase transition from unstable (non-physical) system to stable (physical) one.  For small  divergence  a phase
transition from large  black holes to small  ones takes place and  for large  divergence  a phase
transition occurs from smaller black holes to larger ones.

\section{ Phase transition of 4D EGB Massive black hole }\label{s5}
In the extended phase space, the mass of the black hole treated as the enthalpy rather then the internal energy of the system.
The cosmological constant is treated as the thermodynamics pressure  $(\Lambda=-8\pi P)$ \cite{08,09} and   corresponding conjugate quantity is volume which is given by
\be
\tilde V_+=\frac{4}{3}\pi\tilde  r_+^3.
\ee
The equations of state (EOS) can obtained by relation (\ref{me}) and  Hawking temperature  as
\be
\tilde P_+=\frac{\tilde T_+}{2 \tilde r_+}+ \frac{\tilde T_+\alpha}{\tilde r_+^3}+\frac{\tilde \alpha}{8\pi \tilde r_+^4}-\frac{1}{8\pi \tilde r_+^2}-\frac{ m^2}{8\pi \tilde r_+^2}\left(c\tilde c_1\tilde  r+c^2c_2\right),\qquad\qquad \tilde v=2\tilde r_+,
\ee
where $\tilde v$ represents specific volume. Using the properties of the inflection points we can encode the information about the EoS in the critical points. The critical point  occurs when $\tilde P_+$ has an inflection point
\be
\frac{\partial\tilde  P_+}{\partial \tilde v}=\frac{\partial^2 \tilde P_+}{\partial\tilde v^2}=0.
\ee
 The corresponding critical point is determined by the following equation:
\be
 \tilde r_+^4-12 \tilde \alpha (\tilde r_+^2+ \tilde \alpha)+ m^2 \tilde r_+^2(c^2c_2 \tilde  r_+^2-6\tilde \alpha(c^2c_2+c\tilde c_1 \tilde r_+))=0.
\label{cricr}
\ee
\begin{table}[ht]
 \begin{center}
 \begin{tabular}{ l | l   | l   | l   | l   }
\hline
            \hline
  \multicolumn{1}{c|}{ $m$} &\multicolumn{1}{c}{$\tilde r_C$}  &\multicolumn{1}{|c|}{$\tilde T_C$}  &\multicolumn{1}{c|}{$\tilde P_C$} &\multicolumn{1}{c}{$\frac{\tilde P_C\,\tilde v_C}{\tilde T_C}$}\\
            \hline
            \,\,\,\,\,0 ~~  &~~1.1370~~  & ~~0.0808~~ & ~~0.0126~ & ~~0.3546~~ \\    
            \,\,\,\,\,1~~ &~~0.8525~~ & ~~0.1327~~ & ~~0.0440~~ &  ~~0.5653~~      \\
            \,\,\,\,\,2~~ &~~0.6714~~  & ~~0.3267~~ & ~~0.1666~~ &  ~~0.6847~~    \\
            \,\,\,\,\,3~~ &~~0.6050~~ & ~~0.6770~~ & ~~0.3920~~ &    ~~0.7006~~  \\
            \,\,\,\,\,4~~ &~~0.5757~~  & ~~1.1598~~ & ~~0.7164~~ &    ~~0.7112~~   \\
            \,\,\,\,\,5~~ &~~0.5606~~  & ~~1.7913~~ & ~~1.1373~~ &    ~~0.7118~~    \\
            \hline 
  \multicolumn{1}{c|}{ $\tilde \alpha$} &\multicolumn{1}{c}{}  &\multicolumn{1}{c}{}  &\multicolumn{1}{c|}{} &\multicolumn{1}{|c}{}\\
            \hline
            \,\,\,\,\,0.1 ~~  &~~0.8525~~  & ~~0.1327~~ & ~~0.0440~~ & ~~0.5653~~ \\            
            \,\,\,\,\,0.2~~ &~~1.1403~~ & ~~0.0214~~ & ~~0.0251~~ &  ~~0.7028~~      \\
            \,\,\,\,\,0.3~~ &~~1.3397~~  & ~~0.0590~~ & ~~0.0184~~ &  ~~0.8356~~    \\
            \,\,\,\,\,0.4~~ &~~1.4951~~ & ~~0.0459~~ & ~~0.0150~~ &    ~~0.9771~  \\
            \,\,\,\,\,0.5~~ &~~1.6231~~  & ~~0.0370~~ & ~~0.0129~~ &    ~~1.1317~~   \\
            \hline 
\hline
        \end{tabular}
        \caption{The table for critical temperature $\tilde T_C$, critical pressure $\tilde P_C$ and ${\tilde P_C\,\tilde v_C}/{\tilde T_C}$ corresponding different value of  $m$ with fixed value of $c=1, \tilde c_1=-0.75, c_2=0.75$ and  $\tilde\alpha=0.1$.}
\label{tr10}
    \end{center}
\end{table}
We find that Eq. (\ref{cricr}) can not be solve analytically. In order to estimate the critical points, namely,  critical horizon radius ($\tilde r_C$), critical temperature ($\tilde T_+$) and critical pressure $(\tilde P_+)$   the numerical solutions of   Eq. (\ref{cricr}) with variation of massive parameter and and GB coupling   are tabulated in the Table (II) and Table (III). It is interesting to see that the  critical pressure and  critical temperature  increase  with increasing  critical radius and  massive parameter  at $\tilde \alpha=0.1$.  However, the  critical pressure and  critical temperature  decrease  with increasing critical radius and GB coupling parameter  for $ m=1$. 
\begin{figure}[ht]
\begin{tabular}{c c c c}
\includegraphics[width=.50\linewidth]{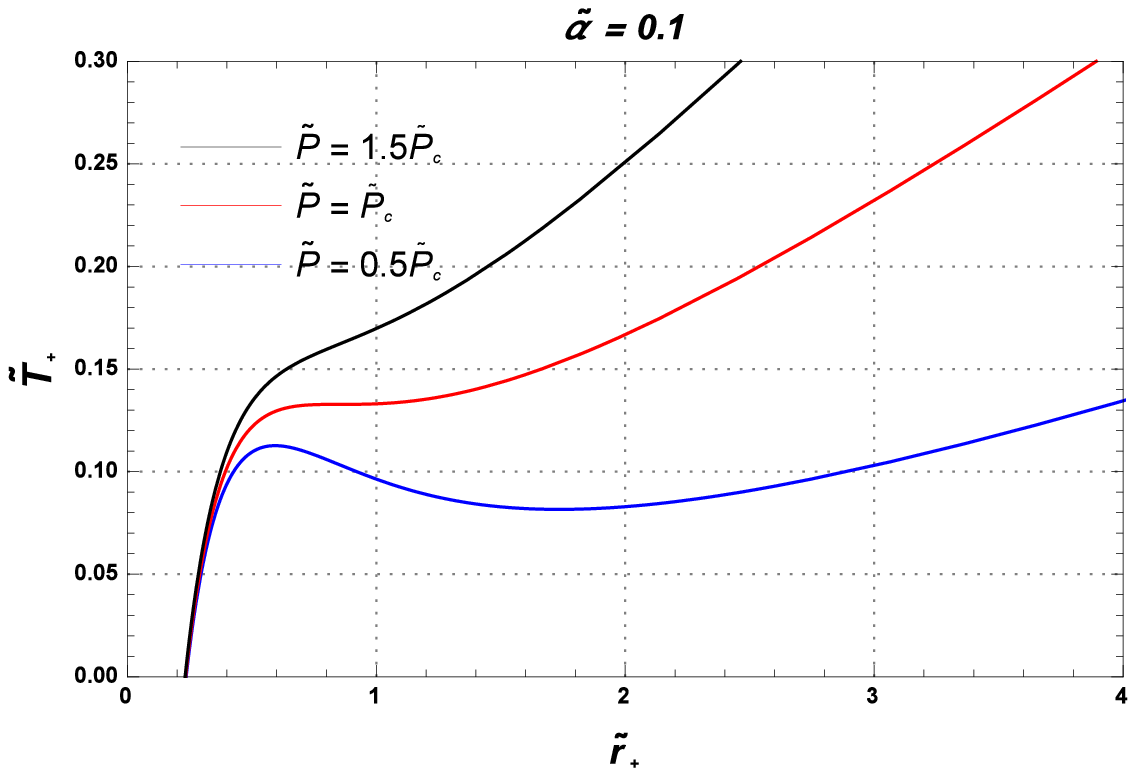}
\includegraphics[width=.50\linewidth]{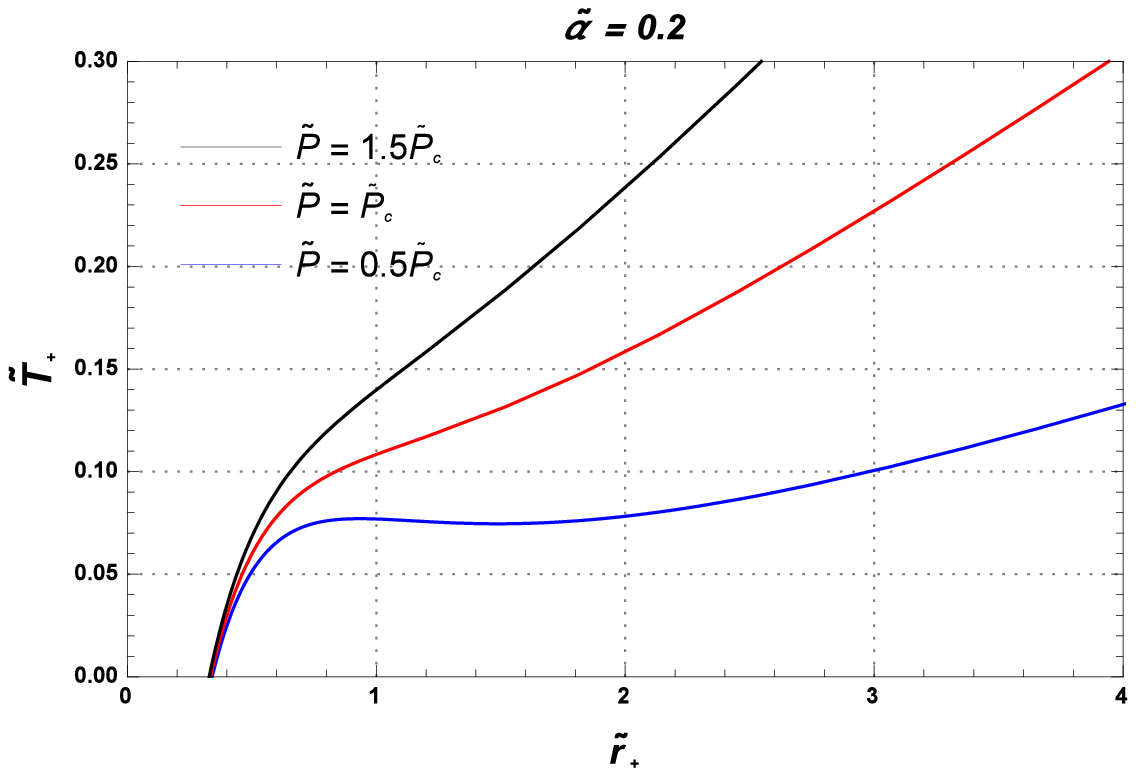}
\end{tabular}
\caption{The plots of temperature  vs horizon radius   for $\tilde \alpha=0.1$ (left) and $\tilde \alpha=0.2$ (right) with fixed values of $c=1,\tilde c_1=-1,c_2=1$ and $m=1$ .}
\label{fig5}
\end{figure}

\begin{figure}[ht]
\begin{tabular}{c c c c}
\includegraphics[width=.50\linewidth]{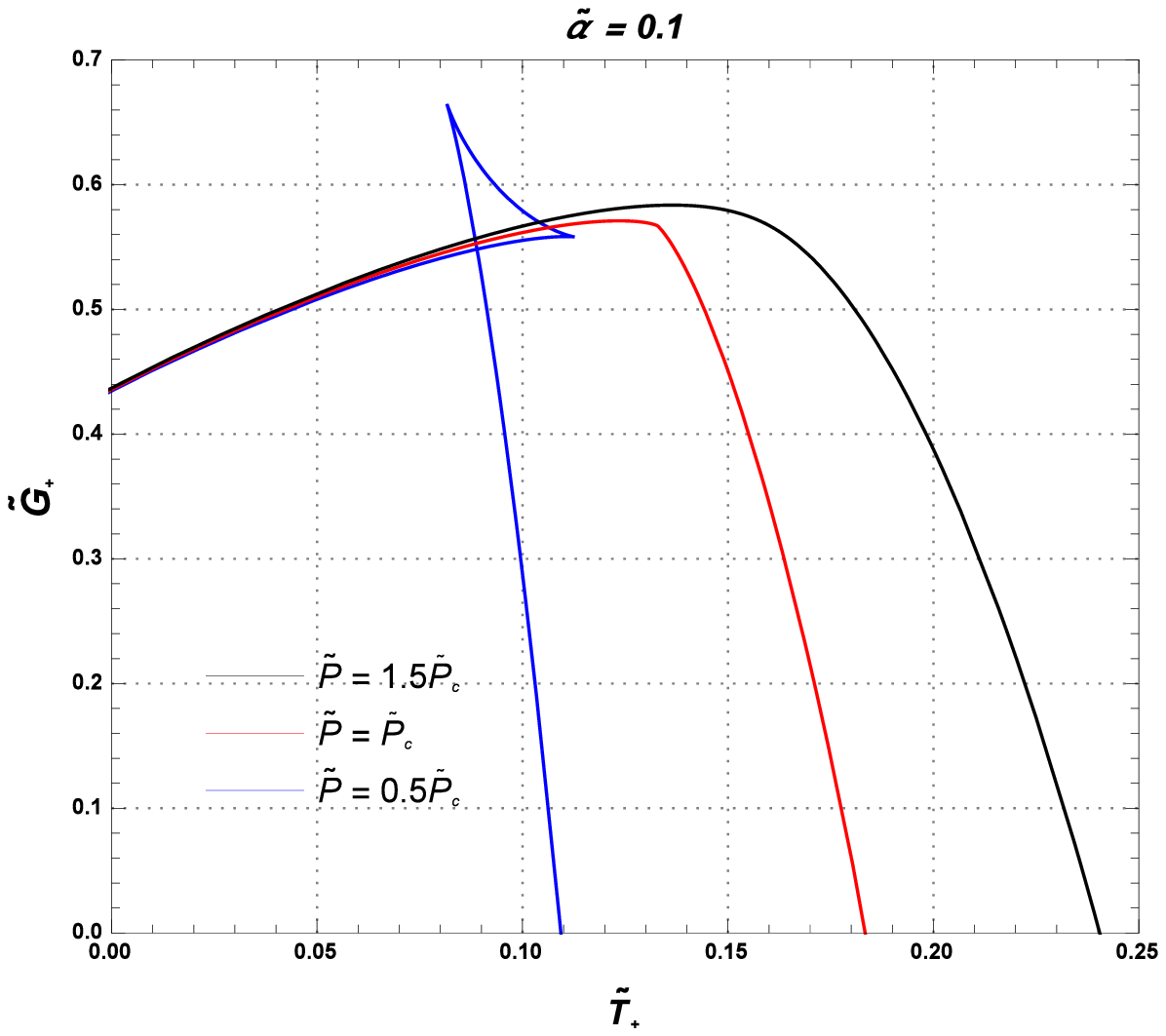}
\includegraphics[width=.50\linewidth]{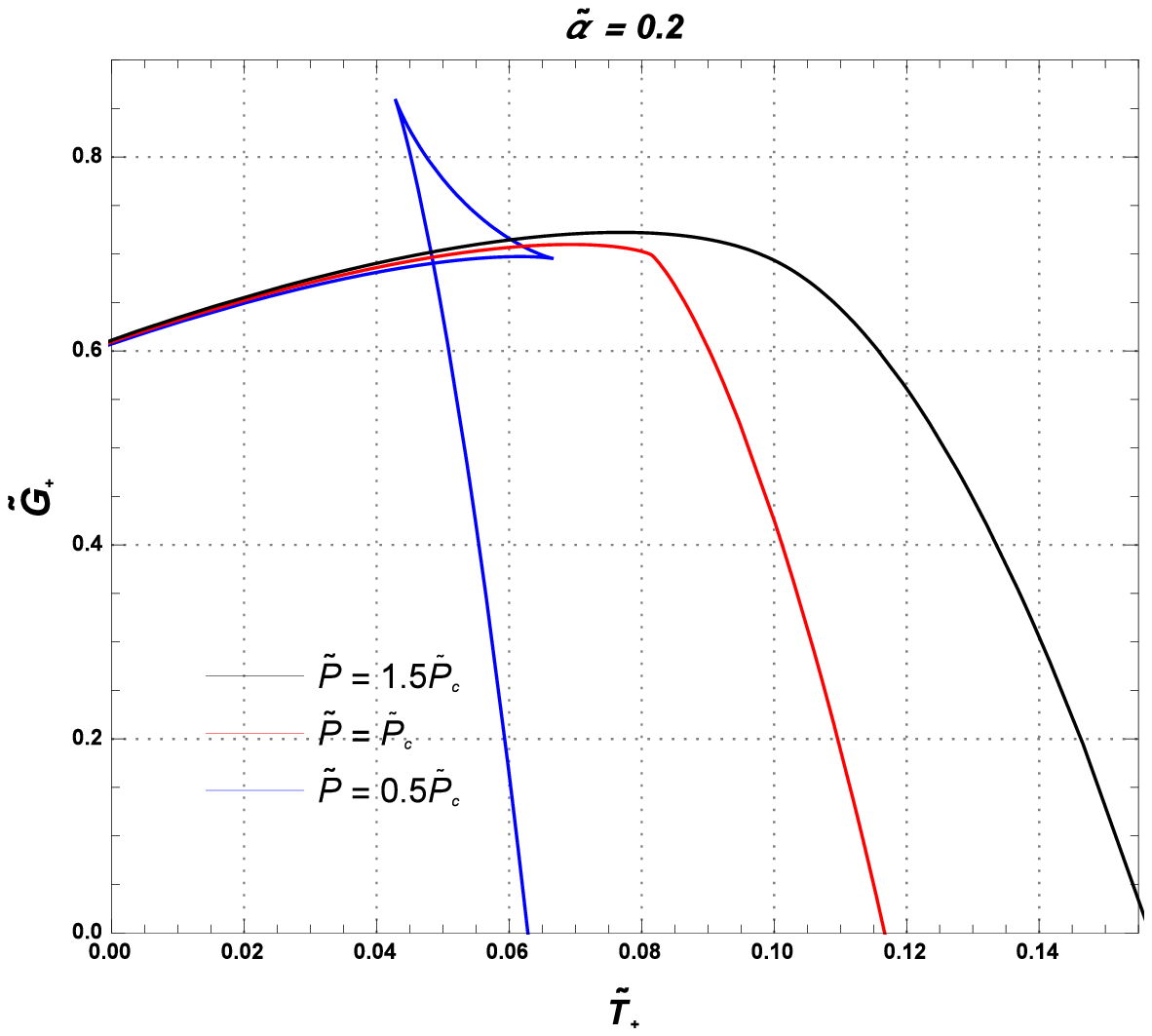}\\
\includegraphics[width=.50\linewidth]{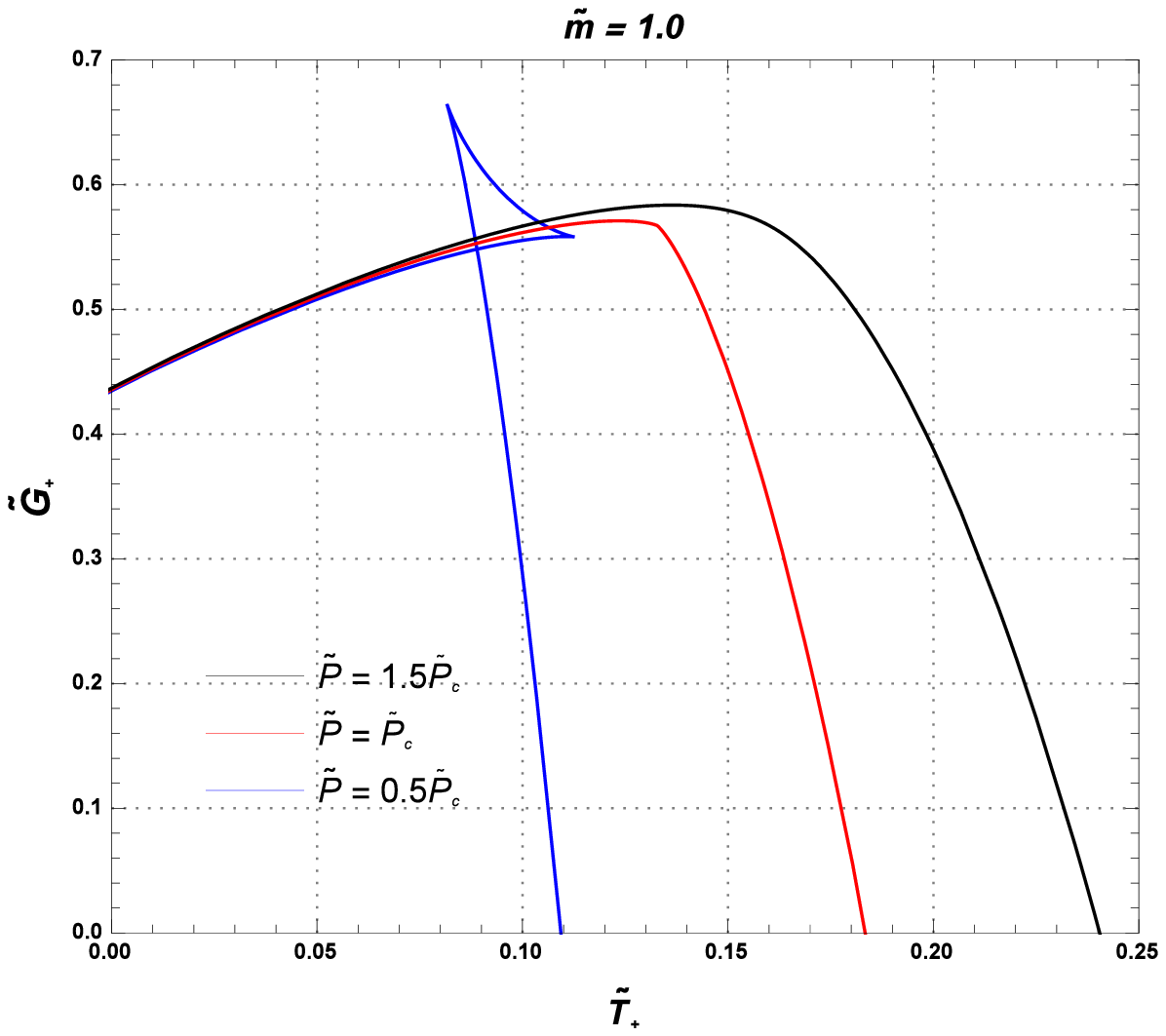}
\includegraphics[width=.50\linewidth]{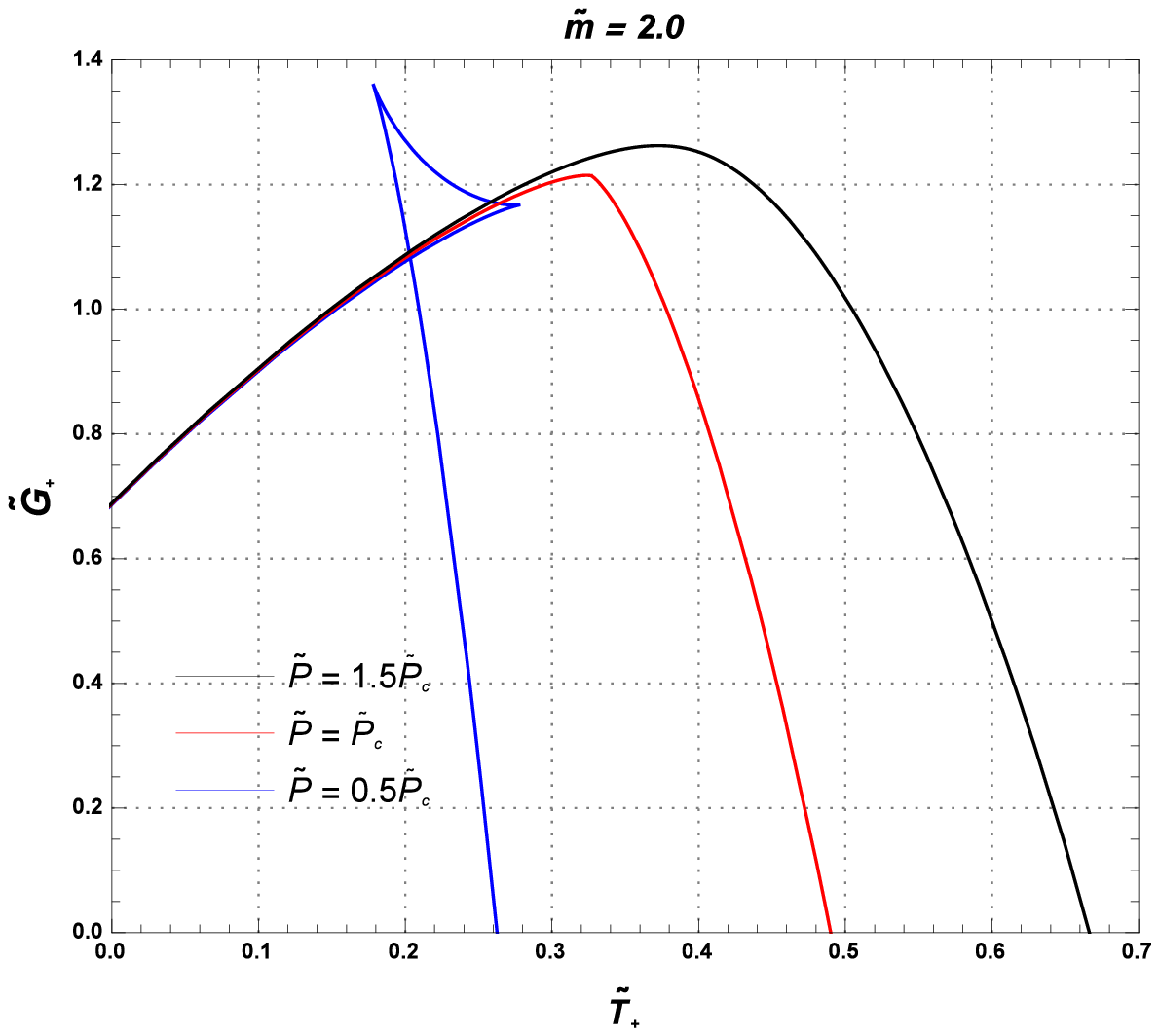}
\end{tabular}
\caption{The plots of  Gibbs free energy ($\tilde G_+$) vs temperature ($\tilde T_+$)  for  $\alpha=0.1$ (upper left), $\alpha=0.2$ (upper right), $ m=1.0$ (lower left) and $ m=2.0$ (lower right) with fixed values of $c=1, \tilde c_1=-0.7, c_2=0.75$ .}
\label{fig7}
\end{figure}

\begin{figure}[ht]
\begin{tabular}{c c  }
\includegraphics[width=.50\linewidth, height=6cm]{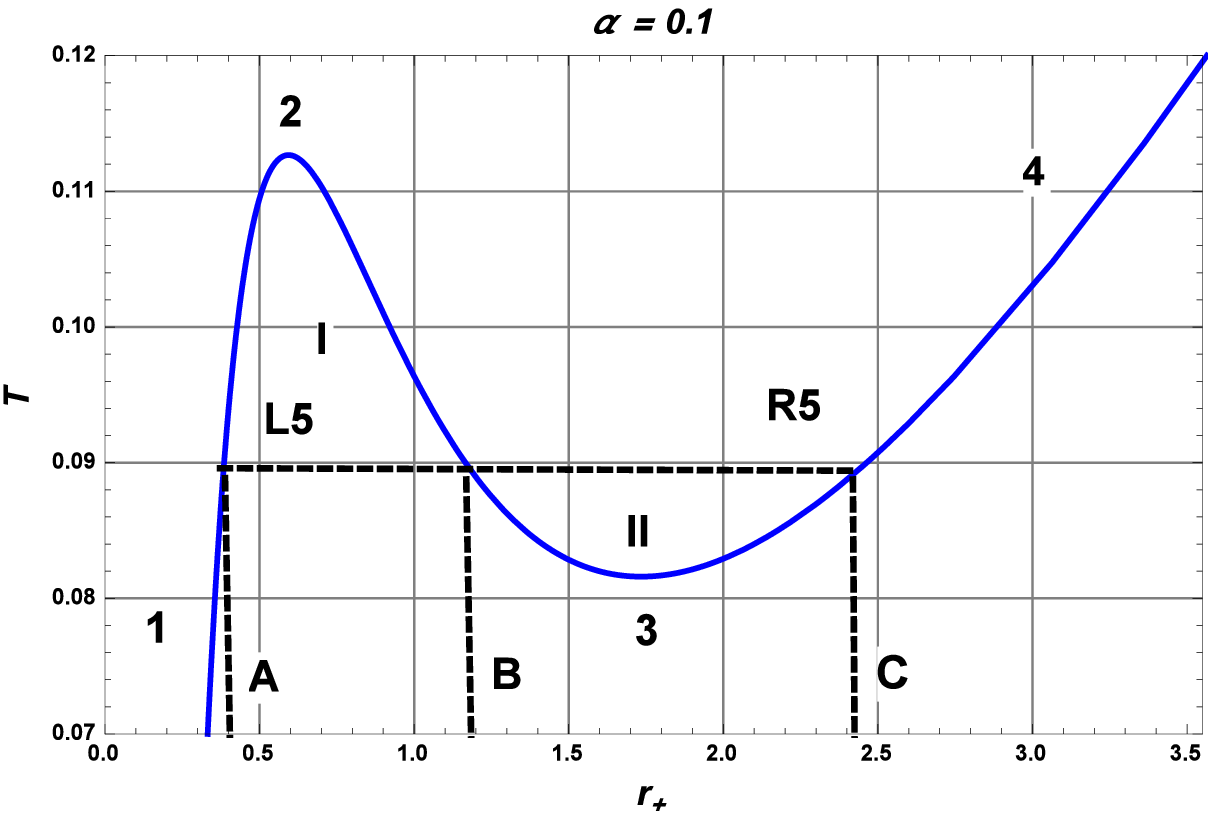}
\includegraphics[width=.50\linewidth, height=6cm]{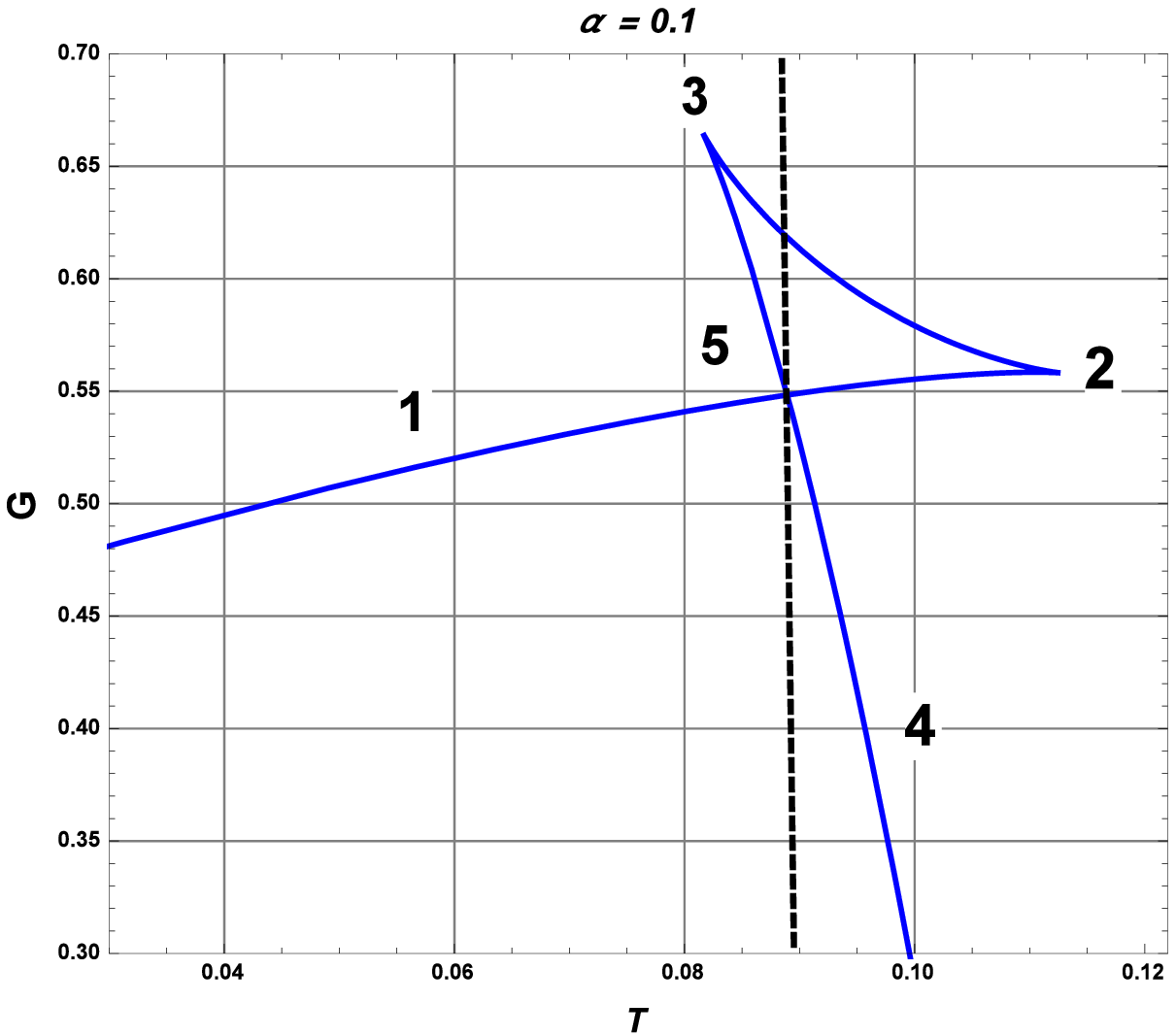}
\end{tabular}
\caption{The plots of  temperature ($\tilde T_+$)  vs horizon radius $(\tilde r_+)$ and   Gibbs free energy ($\tilde G_+$) vs temperature ($\tilde T_+$)  with fixed values of $\tilde \alpha=0.1$ and for  $ m=1.0$ and $ m=2.0$ $c=1, \tilde c_1=-1, c_2=1$.}
\label{fig09}
\end{figure}

 We study the phase transition by analyzing the Gibbs free energy  for different values of  massive gravity parameter $(m)$ and   GB coupling $(\alpha)$  as depicted from    Fig. \ref{fig7}. The characteristic swallow tail   confirms  that the obtained values are the critical ones in which the phase transition takes place. The thermodynamically stable (unstable) state occurs corresponding to the lowest (highest) Gibbs free energy. As a result, the triangular loop in the $\tilde G_+-\tilde  T_+$ plane represents an unstable state. The straight line corresponds to the first order phase transition as the temperature rises in the system, as shown in the $\tilde T_+-\tilde r_+$ plane (Fig. \ref{fig09}). The curved portion of the isobar indicates the unstable state due to the system's possessing lower Gibbs free energy. We note  that the net change in Gibbs free energy around the triangular loop is zero. 

We can also see that swallow tail shape occurs in case of $\tilde P_+<\tilde P_c$  for the first order phase transition.  However, in case of $\tilde P_+=\tilde P_c$, the second order phase transition occurs. The intercepts of the axis in the $\tilde G_+-\tilde T_+$ plane decrease as the GB coupling parameter and mass gravity parameter increase, whereas the intercepts of the $\tilde G$ -axis increase as the GB coupling parameter and mass gravity parameter increase.

 When the temperature is lower than the critical temperature in the $\tilde T_+-\tilde r_+$ plot, we can see three black holes (small, intermediate, and large) for the $4D  AdS$ EGB massive black hole with the same massive gravity parameter of $m$ for a specific temperature range. The small and large black holes are stable, but the intermediate black hole is unstable, since the heat capacity $\tilde C_+$ is negative (see  Fig. \ref{fig4}). Because of the small free energy,   $ \tilde T_+<T_{\star} $ corresponds to a small black hole, and  $ \tilde T_+>T_{\star} $  corresponds to a large black hole, where $T_{\star}$ is the transition temperature, which has values of $0.088$ and $0.048$ for $\tilde\alpha=0.1$ and $\tilde \alpha=0.2$, respectively. Due to the same free energy, black holes can transit from one phase to another phase at a critical temperature. In   Fig. \ref{fig09},   point 5 represents the coexistence of small and large black holes. The small black holes are represented by the solid lines $1-L5$ or $1-5$, while the large black holes are represented by the solid lines $R5-4$ or $5-4$.  We noticed that the point $L5$ and $R5$ share the same free energy. 

\section{Curvature Singularity and Phase Transition  new thermodynamic geometry }
we have proposed a new formalism of the thermodynamic geometry (NTG) which explains the one-to-one correspondence between phase transitions and singularities of the scalar curvature \cite{HosseiniMansoori:2019jcs,HosseiniMansoori:2020yfj}.The NTG geometry is defined as
\begin{equation}
dl^2_{NTG}=\frac{1}{T}\left(n^j_i\frac{\partial^2\Xi}{\partial X^i\partial X^j}dX^idX^j\right),
\label{NTG}
\end{equation}
where $(n^j_i=\text{diag}(-1,1,1,\hdots 1)$
 and $Xi$ is thermodynamic potential \cite{HosseiniMansoori:2019jcs} and the geometrothermodynamics (GTD) metric is conformally related to NTG metric such that this conformal transformation is singular at unphysical points were generated in GTD metric \cite{HosseiniMansoori:2019jcs}.

To study the curvature singularity and phase transition in NTG. We write the  free energy in term  of specific volume and temperature 
\bea
\tilde G_+&=&\frac{1}{\tilde v_+}\left[\frac{\tilde v_+^2}{4}+2\tilde \alpha+\frac{\tilde v_+^3}{8}+m^2\left(\frac{c \tilde c_1\tilde  v_+^2}{8}+\frac{c^2c_2 \tilde v_+}{2}\right)-(\frac{\tilde v_+^2}{4}+4\alpha \log[\frac{\tilde r}{2}]) \tilde T\right].\label{gibbs1}
\eea

The differential form of free energy $d\tilde F_+ =-\tilde S_+d\tilde T_+-\tilde Pd\tilde V_++\tilde Ad\tilde \alpha+\tilde {\cal C}d\tilde c_1$ combining with  Eq. (\ref{gibbs1}), we deduce
\begin{eqnarray}
&&\tilde S_+=-\left(\frac{\partial \tilde F_+}{\partial \tilde T_+}\right)_{\tilde V,\tilde \alpha, {\cal \tilde C}}=\frac{\pi}{4}\left[\tilde v^2+16\tilde \alpha \log\left( \frac{\tilde v}{\tilde v_0}\right)\right],\\
&&\tilde  P=-\left(\frac{\partial \tilde F_+}{\partial \tilde V_+}\right)_{\tilde V,\tilde \alpha, {\cal \tilde C}}=\frac{\tilde T_+}{\tilde v_+}+ \frac{8\tilde v_+\alpha}{\tilde r_+^3}+\frac{2 \tilde \alpha}{\pi \tilde v_+^4}-\frac{1}{2\pi \tilde v_+^2}-\frac{ m^2}{2\pi \tilde v_+^2}\left(\frac{c\tilde c_1\tilde  v_+}{2}+c^2c_2\right),\\
&&\tilde  A=-\left(\frac{\partial \tilde  F_+}{\partial \tilde \alpha}\right)_{\tilde V,\tilde \alpha, {\cal \tilde C}}=\frac{1}{\tilde v}-4\pi \tilde T_+ \log\left( \frac{\tilde v}{\tilde v_0}\right),\\
&&{\cal \tilde C}=-\left(\frac{\partial \tilde F}{\partial \tilde  c_1}\right)_{\tilde V,\tilde\alpha, {\cal \tilde C}}=\frac{ m^2 \tilde c_1 \tilde v^2}{16}.
\end{eqnarray}

The phase transition is appear when a heat capacity diverges, when $\tilde C_+> 0$ the system is stable and $\tilde C_+<0$ the system is unstable. The heat capacity is written as 
\be
\tilde C_+=\frac{{2  \pi \tilde r^2 (\tilde r^2 + 2\tilde \alpha)^2}\left (1 + c^2 c_2  m^2 + c\tilde c_1  m^2 r + {3\tilde r^2} - \frac{\tilde \alpha}{\tilde r^2}\right)}{3 (\tilde r^6 + 6 \tilde r^4 \tilde \alpha) + 
  [-(1 + c^2 c_2  m^2)\tilde r^4 + (5 + 2 c^2 c_2  m^2)\tilde r^2\tilde \alpha + 
    4 c\tilde c_1 m^2 \tilde r^3 \tilde \alpha + 2 \tilde \alpha^2]}.
\label{sh2}
\ee
Now, we are able to implement the NTG geometry to analysis the phase transition behaviour of heat capacity (\ref{sh2}). By substituting the $\Xi=\tilde H=\tilde M_++\tilde E+\tilde P\tilde V_+$ with $X^i=(\tilde S_+,\tilde P, \tilde \alpha,{\cal \tilde C})$ into Eq. (\ref{NTG}), we have 

\[g_H^{NTG}=
  \begin{bmatrix}
    -\left(\frac{\partial \tilde T_+}{\partial \tilde S_+}\right)_{\tilde P,\tilde c_1,\tilde \alpha} &0 &0 & 0 \\
    0 & \left(\frac{\partial \tilde V_+}{\partial \tilde S_+}\right)_{\tilde S_+,\tilde c_1,\tilde \alpha} &\left(\frac{\partial \tilde V_+}{\partial \tilde c_1}\right)_{\tilde S_+,\tilde P,\tilde \alpha}& \left(\frac{\partial \tilde A}{\partial \tilde P}\right)_{\tilde S_+,\tilde c_1,\alpha} \\
    0 &\left(\frac{\partial \tilde V_+}{\partial \tilde C}\right)_{\tilde S_+,\tilde P,\tilde\alpha} & \left(\frac{\partial \tilde C}{\partial \tilde c_1}\right)_{\tilde S_+,\tilde P,\tilde \alpha}& \left(\frac{\partial \tilde A}{\partial \tilde c_1}\right)_{\tilde S_+,\tilde P,\tilde \alpha} \\
    0 & \left(\frac{\partial \tilde A}{\partial \tilde P}\right)_{\tilde S_+,\tilde c_1,\tilde \alpha}  & \left(\frac{\partial \tilde A}{\partial \tilde c_1}\right)_{\tilde S_+,\tilde P,\tilde\alpha}  & \left(\frac{\partial \tilde A}{\partial \tilde \alpha}\right)_{\tilde S_+,\tilde c_1,\tilde P} 
  \end{bmatrix}
\].

Since all thermodynamic parameters are written as a function of $(\tilde  T_+, \tilde v, \tilde c_1, \tilde \alpha)$, it is convenient
to recast metric elements from the coordinate $X_i = (\tilde S, \tilde P, \tilde c_1, \tilde \alpha)$ to the coordinate
$(\tilde T_+,\tilde v,\tilde  c_1,\tilde \alpha)$ by using the following Jacobians

 \[J=\frac{\partial (\tilde S_+,\tilde P, {\cal \tilde C},\tilde \alpha)}{(\tilde T_+, \tilde v,{\cal \tilde C},\tilde \alpha)}=
  \begin{bmatrix}
0&\pi\left(\frac{ \tilde v}{2}+\frac{4\tilde\alpha}{ \tilde v}\right) &0 & {4\pi \log\left( \frac{\tilde v}{\tilde v_0}\right)} \\
    \frac{ \tilde v^2+8 \tilde \alpha}{\tilde  v^3} & \frac{\tilde v^2-\pi \tilde T_+\tilde  v^3-8\tilde(1+3\pi \tilde T \tilde v)}{\pi \tilde  v^5}+\frac{ m^2(4c^2c_2+c \tilde c_1 \tilde v))}{4\pi \tilde v^3}&-\frac{ m^2}{2\pi \tilde  v_+^2}\left(\frac{\tilde c_1\tilde v}{2}+2cc_2\right)&\frac{2(1+4\pi \tilde T\tilde v)}{\pi \tilde v^4} \\
    0 &0 & 1& 0 \\
    0 & 0  & 0&1
  \end{bmatrix}
\]
Finally the metric  element convert to
\[{\hat g}=J^Tg_H^{NTG}J=
  \begin{bmatrix}
0 &0 &0 & -\frac{4\pi}{\tilde T_+} \log\left( \frac{\tilde v}{\tilde v_0}\right)\\
    0 & \frac{\tilde v^2-\pi \tilde T_+\tilde  v^3-8\tilde(1+3\pi \tilde T_+ \tilde v)}{\pi \tilde T_+ \tilde v^3}+\frac{ m^2(4c^2c_2+c \tilde c_1 \tilde v))}{4\pi \tilde  T_+ \tilde v}&0& 0 \\
    0 &0 & \frac{2}{\tilde T_+ \tilde v}& 0 \\
   -\frac{4\pi}{\tilde T_+} \log\left( \frac{\tilde v}{\tilde v_0}\right) &0  & 0&0
  \end{bmatrix}
\].
The denominator of the scalar curvature is
\begin{equation}
D(R^{NTG})=\pi \left(\frac{\tilde v^2-\pi \tilde T_+ \tilde v^3-8(1+3\pi \tilde T_+ \tilde v)}{\pi  \tilde v^5}+\frac{ m^2(4c^2c_2+c\tilde c_1 \tilde v))}{4\pi \tilde v^3}\right)^2\log\left( \frac{\tilde v}{\tilde v_0}\right)
\end{equation}

\begin{figure}[ht]
\begin{tabular}{c c c c}
\includegraphics[width=.50\linewidth]{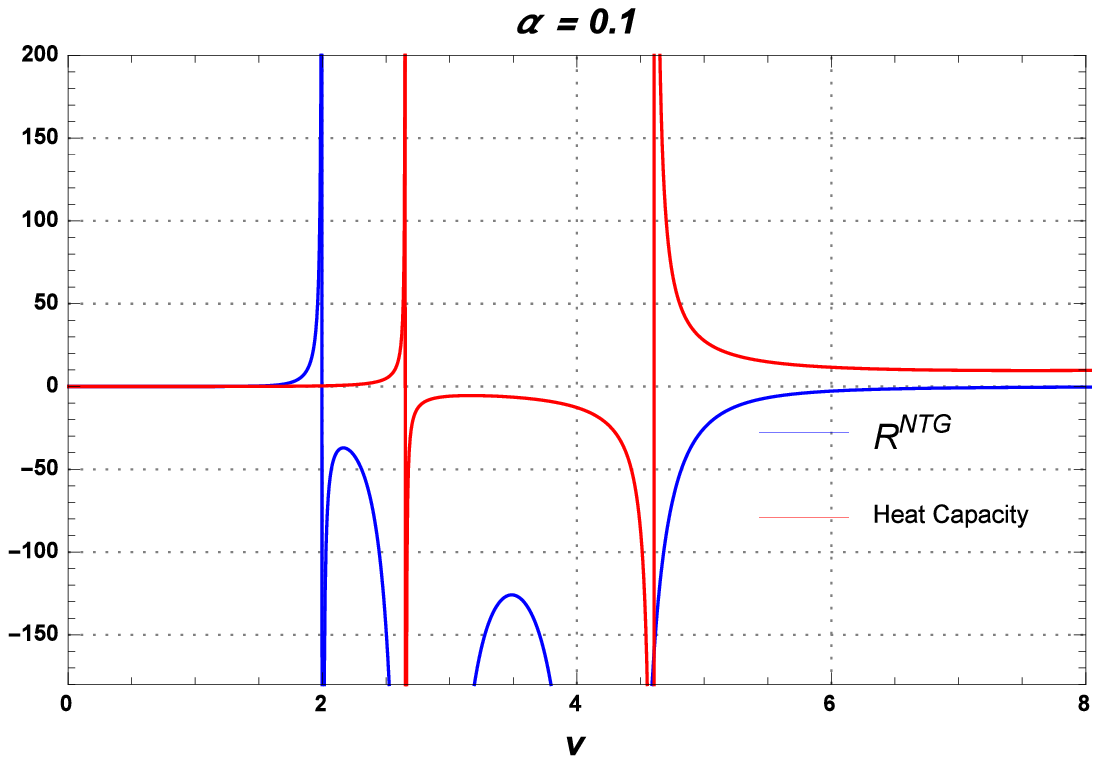}
\includegraphics[width=.50\linewidth]{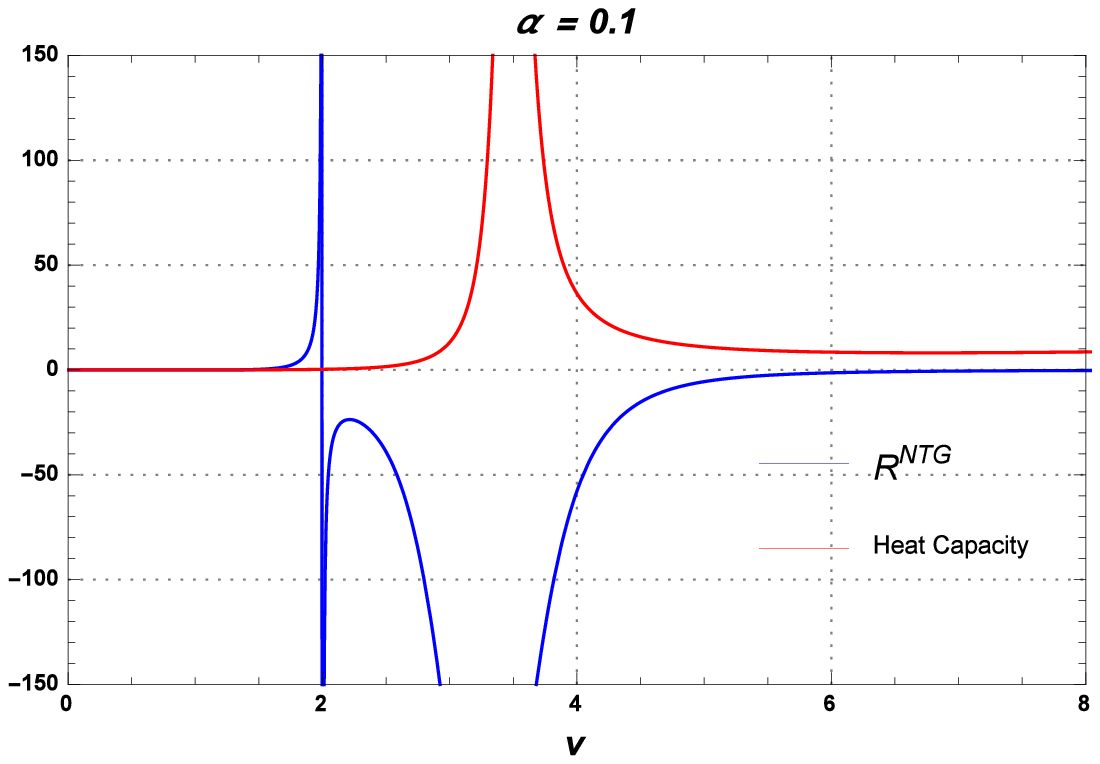}\\
\includegraphics[width=.50\linewidth]{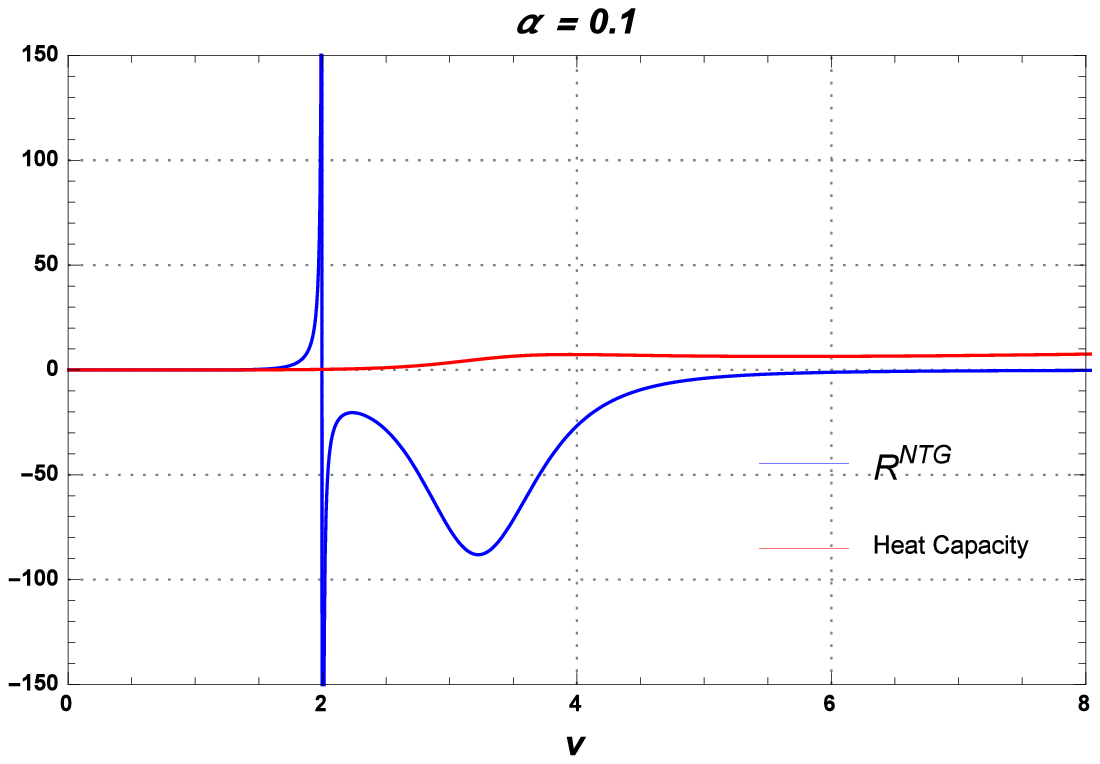}
\end{tabular}
\caption{The plots of  heat capacity and scalar curvature vs specific volume for  $\tilde \alpha=0.1$, $m=2.0$ with fixed values of $c=1,\tilde  c_1=-1.0, c_2=1.0$ for $\tilde T_+ < \tilde T_c, \tilde T_+ = \tilde T_c$ and $\tilde T_+ > \tilde T_c$, where $\tilde T_c=0.3267$.}
\label{fig8}
\end{figure}
The plot of  scalar curvature and heat capacity  with respect to specific
volume are depicted  Fig. \ref{fig8}. We observe that from the  Fig. \ref{fig8}, there is two divergence for heat capacity at $\tilde T_+ < \tilde T_c$ and there are three possible phases  i.e., the small black hole (SBH), intermediate black hole (IBH) and large black hole (LBH). At $\tilde T_+=\tilde T_c$, these divergent points get closer and coincide to form a single divergent point. We noticed that at $\tilde T_+>\tilde T_c$ the heat capacity is positive and  there is no phase transition, it means that the black hole is stable. The scalar
curvature is positive in the range of $0 <\tilde v < \tilde v_0 = 1.7$  that implies a repulsive interaction between the microscopic black hole molecules. 
\section{Conclusions}\label{s6}
We have found a singular solution to the massive EGB gravity in $4D$   $AdS$ spacetime. In fact, we obtain both positive and negative branches of solutions in which one is not physically correct and therefore can be dropped. The horizon structure of the black hole is also discussed. We have found that the real roots of the physically correct metric lead to three horizons (Cauchy, event, and cosmological horizons) below a certain graviton mass limit. However, only two horizons (event and cosmological horizons) are possible after this mass limit.

The thermal properties of this black hole solution are also discussed in a standard way. We derived the mass of the black hole by setting the metric function to zero. The Hawking temperature is calculated from the well-known area-law. We have seen that the black hole solution is not physical when Cauchy and event horizons merge. We have further checked the validity of the first law of black hole chemistry. Gibbs energy is also derived. In order to check the stability of the black hole, we have computed its heat capacity. Here we have one root and two divergent points. Finally, we have described the effects of massive graviton and GB parameters on the phase transition and critical points of the system. We have observed that the critical pressure and critical temperature increase with increasing critical radius and graviton mass. In contrast, the critical pressure and critical temperature decrease as the critical radius and GB parameter increase. In fact,  black holes undergo a phase transition   at a critical temperature. In the Gibbs free energy plots we have seen characteristic swallow tail which locates  the critical point  where phase transition takes place.



\begin{thebibliography}{99}
\bibitem{1}  C. Barrabes and G.F. Bressange,  Class. Quant. Grav. {\bf 14} (1997) 805.
\bibitem{2} R.-G. Cai and Y.S. Myung,  Phys. Rev. D {\bf 56} (1997) 3466.
\bibitem{3} S. Capozziello and A. Troisi,   Phys. Rev. D {\bf 72} (2005) 044022.
\bibitem{4} T.P. Sotiriou,   Class. Quant. Grav. {\bf 23} (2006) 5117.
\bibitem{5} J.W. Moffat,   JCAP {\bf 03} (2006) 004. 
\bibitem{6} V. Faraoni,  Phys. Rev. D {\bf 75} (2007) 067302.
\bibitem{7} K.-i. Maeda and Y. Fujii,   Phys. Rev. D {\bf 79} (2009) 084026.
\bibitem{8} D. Lovelock,  J. Math. Phys. {\bf 12} (1971) 498.
\bibitem{9} D. Lovelock,  J. Math. Phys. {\bf 13} (1972) 874.
\bibitem{10} N. Deruelle and L. Farina-Busto, Phys. Rev. D {\bf 41} (1990) 3696.
\bibitem{11} J.M. Cline and H. Firouzjahi,  Phys. Rev. D {\bf 64} (2001) 023505.
 \bibitem{12} T. Nihei, N. Okada and O. Seto, Phys. Rev. D {\bf 71} (2005) 063535.
 \bibitem{13} M. Demetrian, Gen. Rel. Grav. {\bf 38} (2006) 953.

\bibitem{lan} C. Lanczos, Ann. Math. {\bf 39}, 842 (1938).
\bibitem{cai}R. Cai,  Phys. Rev. D {\bf 65}, 084014 (2002).
\bibitem{har}H. C.D. Lima Junior, C. L.Benone and L. C.B. Crispino, Phys. Lett. B {\bf 811} (2020) 135921.

 
 \bibitem{14}M. Fierz,   Helv. Phys. Acta {\bf 12} (1939) 3. 

 \bibitem{15} D. G. Boulware and S. Deser,  Phys. Rev. D {\bf 6} (1972) 3368.
\bibitem{16}S. H. Hendi,  S. Panahiyan, S. Upadhyay   and B. Eslam Panah,  Phys. Rev. D {\bf 95}, 084036 (2017).
\bibitem{17}S. Upadhyay, S. H. Hendi,  S. Panahiyan   and B. Eslam Panah,  Prog. Theor. Exp. Phys.   093E01, 1-20 (2018).
   
\bibitem{Hendi:2015pda} S.~H.~Hendi, S.~Panahiyan and B.~Eslam Panah, JHEP  {\bf 01}, 129 (2016).
\bibitem{gla}D. Glavan and C. Lin, Phys. Rev. Lett. {\bf 124}, 081301 (2020).
\bibitem{Ao1}
K. Aolki, M. A. Gorji and S. Mukohyama, Phys. Lett. B { 810} (2020) 135843.
\bibitem{Ao2}
K. Aolki, M. A. Gorji and S. Mukohyama, JCAP { 2009} (2020) 014.



\bibitem{fran} P. G. S. Fernandes, Phys. Lett. B  {805} (2020), 135468.
\bibitem{Hennigar:2020lsl}R.~A.~Hennigar, D.~Kubiz\v{n}\'ak, R. B. Mann and C.~Pollack, JHEP {07} (2020) 027.

\bibitem{Singh:2020mty}
S.~G.~Ghosh, D.~V.~Singh, R.~Kumar and S.~D.~Maharaj,
Annals Phys. \textbf{424} (2021), 168347.
\bibitem{Wei:2020poh}
S.~W.~Wei and Y.~X.~Liu,
Phys. Rev. D \textbf{101} (2020) no.10, 104018.
\bibitem{Singh:2020nwo}
D.~V.~Singh, S. G.~Ghosh and S. D. Maharaj, Phys. Dark Univ. {30} (2020) 100730.

\bibitem{Singh:2020xju}
D.~V.~Singh and S.~Siwach,
Phys. Lett. B. {\bf 408} 135658 (2020).

\bibitem{Singh:2021xbk}
D.~V.~Singh, B.~K.~Singh and S.~Upadhyay,
Annals Phys. \textbf{434} (2021), 168642.


\bibitem{Singh21}
B.~K.~Singh, R.~P.~Singh and D.~V.~Singh,
Eur. Phys. J. Plus {136} (2021) no.5, 575.
\bibitem{Singh20}
B.~K.~Singh, R.~P.~Singh and D.~V.~Singh,
Eur. Phys. J. Plus \textbf{135} (2020) no.10, 862.
\bibitem{Godani:2022jwz}
N.~Godani, D.~V.~Singh and G.~C.~Samanta,
Phys. Dark Univ. \textbf{35} (2022), 100952.
\bibitem{hendi19}
S. H. Hendi, M. Momennia, J. High Energ. Phys. 10  (2019) 207.
\bibitem{hendi20}
B. Eslam Panah, Kh. Jafarzade, S. H. Hendi, Nuclear Physics B 961 (2020) 115269.
\bibitem{hendi18}
S. H. Hendi, M. Momennia,  Physics Letters B  777  (2018) 222.
\bibitem{hendi17}
S. H. Hendi, B. Eslam Panah, S. Panahiyan,  M. S. Talezadeh, Eur. Phys. J. C 77 (2017) 133.
\bibitem{hendi16}
S. H. Hendi, B. Eslam Panah, S. Panahiyan ,Class. Quant. Grav. 33 (2016) 
235007.
\bibitem{hendi191}
S. H. Hendi, S. Panahiyan, Phys. Rev. D  90 (2014)   124008.
 
 
\bibitem{hendi2017} 
S.~H.~Hendi, B.~E.~Panah and S.~Panahiyan,
Fortsch. Phys. {66} (2018) no.3, 180000.
 
\bibitem{08} D. Kubiznak and R.B. Mann,  JHEP 07 (2012) 033.
\bibitem{09} S. Gunasekaran, R.B. Mann and D. Kubiznak,   JHEP 11 (2012) 110.
\bibitem{HosseiniMansoori:2019jcs}
S.~A.~Hosseini Mansoori and B.~Mirza,
Phys. Lett. B \textbf{799} (2019), 135040
\bibitem{HosseiniMansoori:2020yfj}
S.~A.~Hosseini Mansoori,
Phys. Dark Univ. \textbf{31} (2021), 100776

\end{thebibliography}
\end{document}